\documentclass[preprint,review, 12pt, 3p]{elsarticle}

\usepackage{amssymb}
\usepackage{amsmath}
\usepackage{subfig}
\usepackage{array,tabularx}
\usepackage{booktabs}
\usepackage[table]{xcolor}
\usepackage{easyReview}

\newcommand{\insitu}[1]{\textit{in situ}#1}
\newcommand{\myetal}[1]{\textit{et al.}#1}
\newcommand{\planewave}[1]{plane-wave#1}
\newcommand{\wavenumber}[1]{wave-number#1}

\newcommand{\reference}[1]{Ref.#1}

\newcommand{\Eq}[1]{Eq.~(\ref{#1})} 
\newcommand{\Equation}[1]{Equation~(\ref{#1})} 
\newcommand{\Figure}[1]{Figure~\ref{#1}}
\newcommand{\Fig}[1]{Fig.~\ref{#1}}

\newcommand{\Tab}[1]{Tab.~(\ref{#1})} 
 
\newcommand{\Sec}[1]{Sec.~\ref{#1}}
\newcommand{\Section}[1]{Section~\ref{#1}}
\newcommand{\eg}{{\it e.g., }} 
\newcommand{\Secs}[1]{Secs.#1}
\newcommand{\Figs}[1]{Figs.#1}
\newcommand{\Eqs}[1]{Eqs.#1}
\newcommand{\Reference}[1]{Ref.#1}
\newcommand{\References}[1]{Refs.#1}
\newcommand{\co}[1]{#1} 
\DeclareMathOperator{\ji}{\hspace{0.2mm}j\hspace{-0.35mm}}
\DeclareMathOperator{\e}{e}
\DeclareMathOperator{\M}{\mathcal{M}}

\DeclareMathOperator{\Q}{\mathcal{Q}}
\newcommand{\vecb}[1]{\textbf{#1}}
\newcommand{\norm}[1]{\left\lVert#1\right\rVert}
\DeclareMathOperator{\argmin}{\underset{s}{\mathrm{argmin}}}
\DeclareMathOperator{\dt}{d\hspace{-0.4mm}}

\newcommand{\txu}[1]{\ensuremath{_{\text{#1}}}} 

\definecolor{LightOrange2}{RGB}{255,235,193} 

\journal{Journal of Sound and Vibration}

\begin{document}

\begin{frontmatter}



\title{In situ sound absorption estimation with the discrete complex image source method}


%
%

\author[inst1,inst2]{Eric Brandão\corref{cor1}} \ead{eric.brandao@eac.ufsm.br}
\author[inst1,inst3]{William D'Andrea Fonseca}
\author[inst1,inst4]{Paulo Henrique Mareze}
\author[inst1]{Carlos Resende}
\author[inst1,inst2]{\\[4pt] Gabriel Azzuz}
\author[inst1]{João Pontalti}
\author[inst5]{Efren Fernandez-Grande}

\affiliation[inst1]{organization={Acoustical Engineering Program, Federal University of Santa Maria (UFSM)}, 
            city={Santa Maria},
            postcode={97105-900}, 
            state={RS},
            country={Brazil}}

\affiliation[inst2]{organization={Graduate Program in Civil Engineering, Federal University of Santa Maria (UFSM)},
            city={Santa Maria},
            postcode={97105-900}, 
            state={RS},
            country={Brazil}}            

\affiliation[inst3]{organization={Graduate Program in Architecture, Urbanism, and Landscaping, Federal University of Santa Maria (UFSM)}, 
            city={Santa Maria},
            postcode={97105-900}, 
            state={RS},
            country={Brazil}}               

\affiliation[inst4]{organization={Graduate Program in Mechanical Engineering, Federal University of Santa Maria (UFSM)},
            city={Santa Maria},
            postcode={97105-900}, 
            state={RS},
            country={Brazil}}           

\affiliation[inst5]{organization={Department of Electrical Engineering, Technical University of Denmark (DTU)},
            city={Kgs. Lyngby},
            postcode={2800}, 
            country={Denmark}}

\cortext[cor1]{Corresponding author}

\begin{abstract}

Estimating the sound absorption \insitu{} relies on accurately describing the measured sound field. Evidence suggests that modeling the reflection of impinging spherical waves is important, especially for compact measurement systems. This article proposes a method for estimating the sound absorption coefficient of a material sample by mapping the sound pressure, measured by a microphone array, to a distribution of monopoles along a line in the complex plane. The proposed method is compared to modeling the sound field as a superposition of two sources (a monopole and an image source). The obtained inverse problems are solved with Tikhonov regularization, with automatic choice of the regularization parameter by the L-curve criterion. The sound absorption measurement is tested with simulations of the sound field above infinite and finite porous absorbers. The approaches are compared to the \planewave{} absorption coefficient and the one obtained by spherical wave incidence. Experimental analysis of two porous samples and one resonant absorber is also carried out \insitu{}. Four arrays were tested with an increasing aperture and number of sensors. It was demonstrated that measurements are feasible even with an array with only a few microphones. The discretization of the integral equation led to a more accurate reconstruction of the sound pressure and particle velocity at the sample's surface. The resulting absorption coefficient agrees with the one obtained for spherical wave incidence, indicating that including more monopoles along the complex line is an essential feature of the sound field.

\end{abstract}


%


\begin{keyword}
Discrete complex image sources \sep Absorption coefficient \sep \textit{In situ} \sep Microphone array



\end{keyword}

\end{frontmatter}

\section{Introduction} \label{sec:intro}


The sound absorption coefficient is one of the most important properties of materials used in noise and reverberation control. The procedures of random- and normal-incidence absorption coefficient measurements are described in the standards ISO~354~\cite{iso3542003} and ISO~105342~\cite{iso1053422001a}, respectively. The \insitu{} or free-field sound absorption measurement is also of considerable interest since it is a non-destructive technique for which it is possible to measure under realistic mounting conditions. Several \insitu{} measurement methods exist, and a review of methods based on a small number of sensors is detailed by Brandão \myetal{}~\cite{Brandao2015}. Most of these techniques are formulated as an inverse problem since the sound pressure and/or particle velocity are measured close to the sample, and the surface impedance and absorption coefficient are estimated from the measurements with a sound field physical model. The complexity and assumptions of such models are known to have an impact on the estimated results. For instance, the impact on the assumption of how spherical waves reflect on infinite locally reacting planar surfaces was investigated by Brandão \myetal{}~\cite{Brandao2011}. In contrast, the effect of local vs. non-local reaction assumption is investigated in \References{}~\cite{Brandao2013, Dragonetti2015, Dragonetti2017}. The quantification of measurement uncertainty in \insitu experiments was also addressed either by Monte-Carlo simulations~\cite{Brandao2011b} or, more recently, by a Bayesian approach introduced by Eser \myetal \cite{eser2023free}. The effects of measuring finite locally-reacting samples were thoroughly documented in relevant research studies \cite{Brandao2012, Hirosawa2009, Brandao2022}. Nevertheless, it is more difficult to find work investigating the measurement of finite and non-locally reacting samples, but experimental reports exist~\cite{Kimura2002, Brandao2011a}.  

The use of microphone arrays to measure sound absorption gained some attention in recent years, following the work of Tamura~\cite{Tamura1990, Tamura1995} in the 1990's. In this case,  a microphone array samples the sound pressure at several positions, and an ill-posed inverse problem is formulated by describing the sound field as a sum of elementary waves. For instance, there are basis functions such as propagating plane-waves~\cite{Richard2019, Nolan2020}, propagating and evanescent plane-waves~\cite{Brandao2022, Hald2019, Ottink2016}, spatially distributed monopoles~\cite{Richard2017}, and spherical harmonics~\cite{Dupont2020}. In addition, different types of arrays can be used, such as spherical~\cite{Richard2017, Dupont2020}, double-layered with regular spacing~\cite{Brandao2022, Hald2019, Ottink2016}, and randomized arrays~\cite{Nolan2020}. Solving the proposed inverse problem requires regularization, typically carried out with Tikhonov regularization or sparse processing~\cite{Richard2017}. The absorption coefficient can be estimated by reconstructing the surface impedance~\cite{Brandao2022, Richard2017} or using the acoustic power of the estimated wave-number spectrum~\cite{Nolan2020}. Also, an array of loudspeakers combined with sound field synthesis was proposed by Dupont \myetal{}~\cite{dupont2023measurement} to measure the diffuse-field absorption coefficient. The use of artificial intelligence in this context was introduced by Zea \myetal{}~\cite{Zea2023} to measure finite and locally reacting samples as if they were infinite.

Alkmin \myetal{}~\cite{alkmim2021angle} investigated the angle-dependent sound absorption estimation using a compact microphone array. One of the proposed inverse problems was formulated considering that the sound field comprised of a monopole (associated with the original sound source) and an image source, similar to the two-microphone method of Li and Hodgson~\cite{Li1997}. The authors stated that their main goal was to improve the estimated absorption coefficient by modifying the classical two-microphone set-up while keeping the measurement apparatus compact. For the more realistic case of spherical wave incidence, the sound field model with a source and an image source is known to be simplistic. For instance, Di and Gilbert~\cite{Di1993} modeled the same problem with an integral equation mapping the complex amplitude of the reflected sound pressure to a distribution of image sources along a complex line. Incorporating this model led to more accurate estimations of sound absorption measured at a single point \insitu{} using a PU probe~\cite{Brandao2011}. 

This article proposes the incorporation of Di and Gilbert's model~\cite{Di1993} in an inverse problem setting, used to estimate the absorption coefficient from distributed sound pressure measurements. The main contribution is the presentation of the formulation herein called the Discrete Complex Image Source Method (DCISM). It is compared to the Image Source Method (ISM) proposed by Alkmin \myetal{}~\cite{alkmim2021angle}. To the best of the author's knowledge, this is the first time that the DCISM has been presented in the literature. The DCISM and the ISM are evaluated experimentally and by simulations of the measurement above infinite and finite non-locally reacting samples. Another contribution of the article is demonstrating that the DCISM can accurately retrieve the absorption coefficient under spherical wave incidence. This is in connection with the work of Dragonetti and Romano~\cite{Dragonetti2015, Dragonetti2017}. Even though the analysis of finite non-locally reacting absorbers is not intended to be extensive, it is rarely explored in the literature, and the aim here is to provide practical guidelines for measurements and future research. Three absorbers were measured: a Polyethylene Terephthalate fibrous absorber (PET), a Melamine foam, and a Helmholtz resonant absorber. The presentation and discussion of the sound absorption of the latter sample is one of the contributions of this article since the \insitu{} measurements of resonant absorbers are rarely found in the literature. Finally, measurements were conducted \insitu{} with the aid of a cost-effective sound field scanner built by the research team. The brief description of such instrumentation is also regarded as a contribution of this research to this field.

The article is structured as follows: in \Sec{sec:matmet}, the methodology for the experimental validation and simulation of the measurements above infinite and finite non-locally reacting samples is described. The instrumentation built and used, the measured samples, and the post-processing strategy are reported for the experiments. An analytical model was used for the simulation of the infinite sample, and a multi-physics Finite/Boundary Element Method was used for the finite sample. \Section{sec:theory} presents the mathematical model inherited in the proposed DCISM and the ISM, the inverse problem estimation strategy, and the sound field reconstruction equations. \Section{sec:results} presents the results and discussions about the simulated and actual experiments, followed by the conclusions in \Sec{sec:conclusions}.

\section{Materials and methods} \label{sec:matmet}


This section describes the simulations and experimental validation conducted for this research. When testing a novel approach to extract the absorption coefficient from measured data, it is common practice to use forward models to simulate the experiment. Herein, this is accomplished both by an analytical and a numerical model, described in \Secs{}~\ref{sec:matmet:inf} and~\ref{sec:matmet:fin}, respectively. The experiments are described in \Sec{sec:matmet:exp}. Common elements between simulations and experiments are discussed in the following.

First, the absorption coefficient of an infinite planar absorber under \planewave{} incidence with an elevation angle $\theta$ is considered. In this scenario, the absorption coefficient is calculated by


\begin{equation}
\label{eq:alpha}
\alpha(\theta) = 1 \ - \ \left|\frac{\; Z\txu{s}\cos{(\theta)}-1 \;}{Z\txu{s}\cos{(\theta)}+1}\right|^{2},
\end{equation}
where $\rho_0$,~kg/m$^3$, and $c_0$,~m/s, are the air density and speed of sound, respectively, and $Z_{\text s}$ is the surface impedance of the sample. Note that $Z_{\text s}$ is constant across the sample's surface for this case. For a layer of thickness $d$ over a rigid backing (considered in this research), the \planewave{} surface impedance is given by

\begin{equation}
\label{eq:zs}
Z_{\text s} = - \ji Z_{\text{p}} \, \frac{\; k_{\text{p}} \;}{k_{\text{pz}}} \cot{\left(k_{\text{pz}} \, d \right)} \,,
\end{equation}
where $\ji=\sqrt{-1}$, and $k_{\text{p}}$ and $k_{\text{pz}}$ are the porous layer complex \wavenumber{} and its component on the $z$-axis, respectively; $Z_{\text{p}}$ is the characteristic impedance of the porous layer. The Johnson-Champoux-Allard (JCA)~\cite{Johnson1987, Champoux1991} model, as in \ref{sec:appendixA}, was used to compute $k_{\text{p}}$ and $Z_{\text{p}}$. Note that \Eq{eq:zs} represents a surface impedance with non-locally reacting behavior, which is also relevant for spherical wave incidence.

\begin{figure}[ht!]
\begin{center}
\includegraphics[width=0.7\columnwidth]{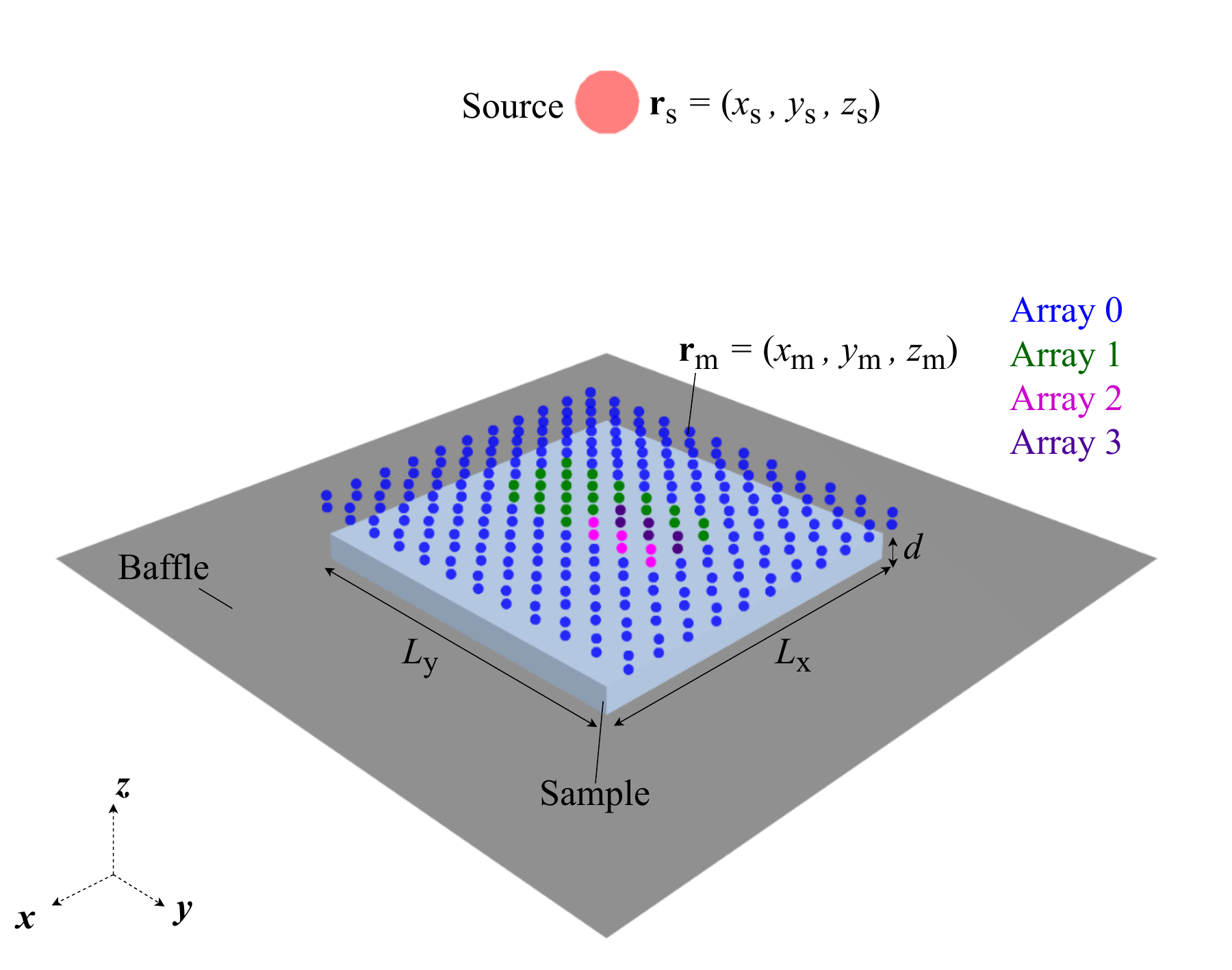}
\caption{\label{fig:schematics}{\co Schematic of the sound scene. The point source (red) at $\vecb{r}_{\text s}$ excites the sound field. The absorber of dimensions $L_x  \times L_y \times d$ (light blue) is baffled at $z=-d$. The Arrays~0--3 are also shown. Note that Array~0 contains Arrays~1--3, Array~1 contains Arrays~2--3, and Array~2 contains Array~3.}}
\end{center}
\end{figure}

\Figure{fig:schematics} shows a schematic drawing of how a typical \insitu{} measurement is conducted. A monopole sound source at coordinate $\vecb{r}_{\text s} = (x_{\text s}, y_{\text s}, z_{\text s})$ creates the sound field above the absorber. In this article, the analysis is restricted to normal incidence, which means that $\vecb{r}_{\text s} = (0, 0, z_{\text s})$. An absorber of thickness $d$~(m) and side lengths $L_{\text{x}} \times L_{\text{y}}$ is placed over a rigid baffle (\eg the floor of the room where measurements were conducted). The coordinate system's origin lies at the sample's center and top.

The forward models in \Secs{}~\ref{sec:matmet:inf} and \ref{sec:matmet:fin} were used to compute the sound field at the receiver points, each at coordinate $\vecb{r}_{\text m} = (x_{\text m}, y_{\text m}, z_{\text m})$. The receivers can be either above the surface of the sample, $z>0$ (\eg an array of microphones used in the experiments, represented by the colored dots in \Fig{fig:schematics}), or directly at the surface of the absorber, $z=0$ (\eg receivers used only in the simulations to establish benchmark results --- see \Fig{fig:scanner_fembem-a}).

For the receivers above the sample's surface, $z>0$, the sound pressure is measured or computed with the forward models at $\M$ points. This data serves as input to the proposed inverse problems in \Sec{sec:theory}. Arrays of different aperture sizes and number of microphones were tested (see \Tab{tab:arrays}), which is relevant because the inherited models in the inverse problems do not account for the edge diffraction on the finite absorbers. This edge diffraction effect is also the reason for conducting the simulations of infinite and finite absorbers. Array~0 (all dots in \Fig{fig:schematics}) has microphones arranged in a double layer of $0.57 \times 0.65$~m$^{2}$ and $11 \times 12$ equally spaced microphones in each layer. The separation between the layers is $2.9$~cm, and the distance between the surface of the sample and the closest layer is $1.3$~cm. The other arrays consist of subsets of Array~0, with smaller apertures and fewer microphones. Thus, Array~0 contains Arrays 1--3, Array~1 contains the receivers in Arrays 2--3, and Array~2 contains Array~3 (a double-layer line array with six receivers). Gaussian noise, with a signal-to-noise ratio (SNR) of 30~dB, was added to all the computed sound pressures in the simulations to emulate noise in the measured scenarios.


\begin{table}[ht!]
	\caption{The Arrays~0--3 apertures, $x$ and $y$ spans and number of microphones.} 
	\centering 
	\begin{tabular}{c | c | c |c |c} 
		\toprule
		 \rowcolor{LightOrange2}  \textbf{Name} & \textbf{Aperture}~cm$^{2}$ &  $x$ \textbf{span}~cm &  $y$ \textbf{span}~cm & $\M$ \textbf{microphones}\\  
  	 \midrule
		\rowcolor[gray]{0.90} Array 0 & $57.00 \times 65.0$ & $\pm 57.0/2$   & $\pm 65.0/2$ & $11 \times 12 \times 2 = 264$  \\ 
		 \rowcolor[gray]{1.00} Array 1 & $11.00 \times 30.0$ & $\pm 5.5$   & $\pm 15.0$ & $3 \times 6 \times 2 = 36$  \\ 
		\rowcolor[gray]{0.90} Array 2 & $5.50 \times 15.0$ & $0.0$ to $5.50$   & $0.0$ to $15.0$ & $2 \times 3 \times 2 = 12$  \\ 
		 \rowcolor[gray]{1.00} Array 3 & $0.00 \times 15.0$ & $0.0$   & $0.0$ to $15.0$ & $1 \times 3 \times 2 = 6$  \\ 
		\bottomrule
	\end{tabular}
	\label{tab:arrays}
\end{table}

\vspace{0.5cm}
For the receivers at the sample's surface, $z =0$, both the sound pressure ($p$) and the $\hat{z}$ component of the particle velocity ($u_{\text{z}}$) were computed. The reason for such simulations is to establish benchmark results to compare with the inverse problem reconstructions of these acoustic quantities (\Secs{}~\ref{sec:ism} and~\ref{sec:dcism}). The sound pressure and particle velocity were calculated for a point at the center of the sample, $(0, 0, 0)$~m, which was used to obtain a point-wise surface impedance for spherical wave incidence (herein referred to as $Z_{\text s} \ @ \ (0, 0, 0)$). It should be noted that this differs from the \planewave{} surface impedance as defined in \Eq{eq:zs}. Additionally, $p$ and $u_{\text{z}}$ were computed at a small patch area of $10 \times 10$~cm$^{2}$ centered at the surface of the sample; the area contains $21 \times 21$ equally spaced points (see \Fig{fig:scanner_fembem-a}). Using such a grid in the \insitu surface impedance measurements is customary~\cite{Brandao2022, Richard2019, Richard2017}, and it allows the computation of the Normalized Mean Square Error (NMSE) of the reconstructed sound field quantities as a function of frequency, which is given by
\begin{equation}
\label{eq:nmse}
\text{NMSE} \,(f) = \frac{\; \left\|\vecb{x}_{\text{re}}-\vecb{x}_{\text{true}}\right\|_{2}^{2} \;}{\left\|\vecb{x}_{\text{true}}\right\|_{2}^{2}} \,,
\end{equation}
%
where $\left\|\cdot\right\|_{2}$ stands for the the $\ell^{2}$-norm of a vector, $\vecb{x}_{\text{true}}$ is the vector containing the true sound pressure or particle velocity computed by the forward model (\Sec{sec:matmet:inf} or \Sec{sec:matmet:fin}) at a given frequency $f$, and $\vecb{x}_{\text{re}}$ is the vector containing either the reconstructed sound pressure or particle velocity.


\subsection{Experiments (specification, hardware, and software)} \label{sec:matmet:exp}

\begin{figure}[ht!]
\centering
\subfloat[]{\label{fig:scanner_fembem-a}
\includegraphics[width=0.40\linewidth]{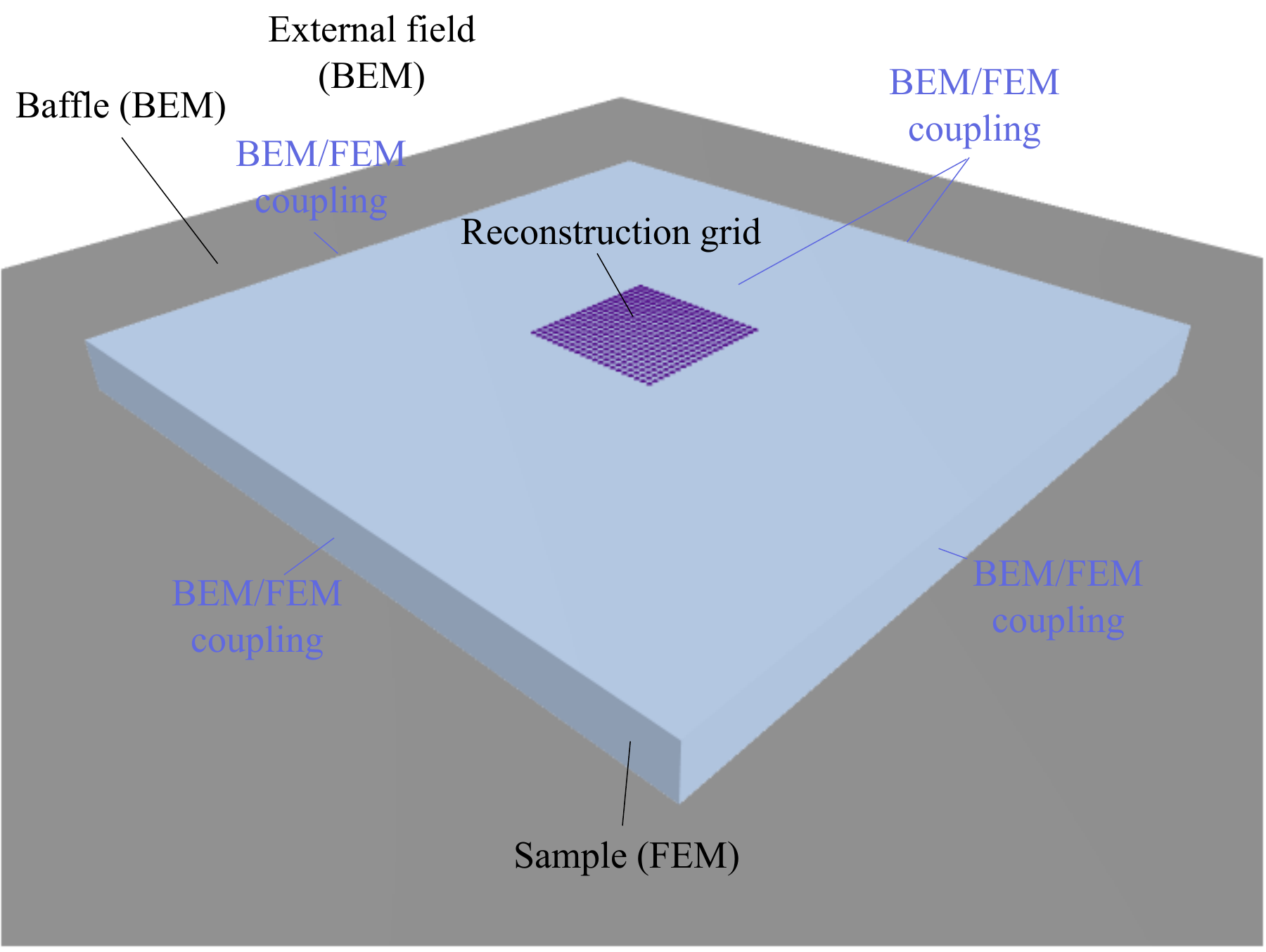}}
\subfloat[]{\label{fig:scanner_fembem-b}
\includegraphics[width=0.48\linewidth]{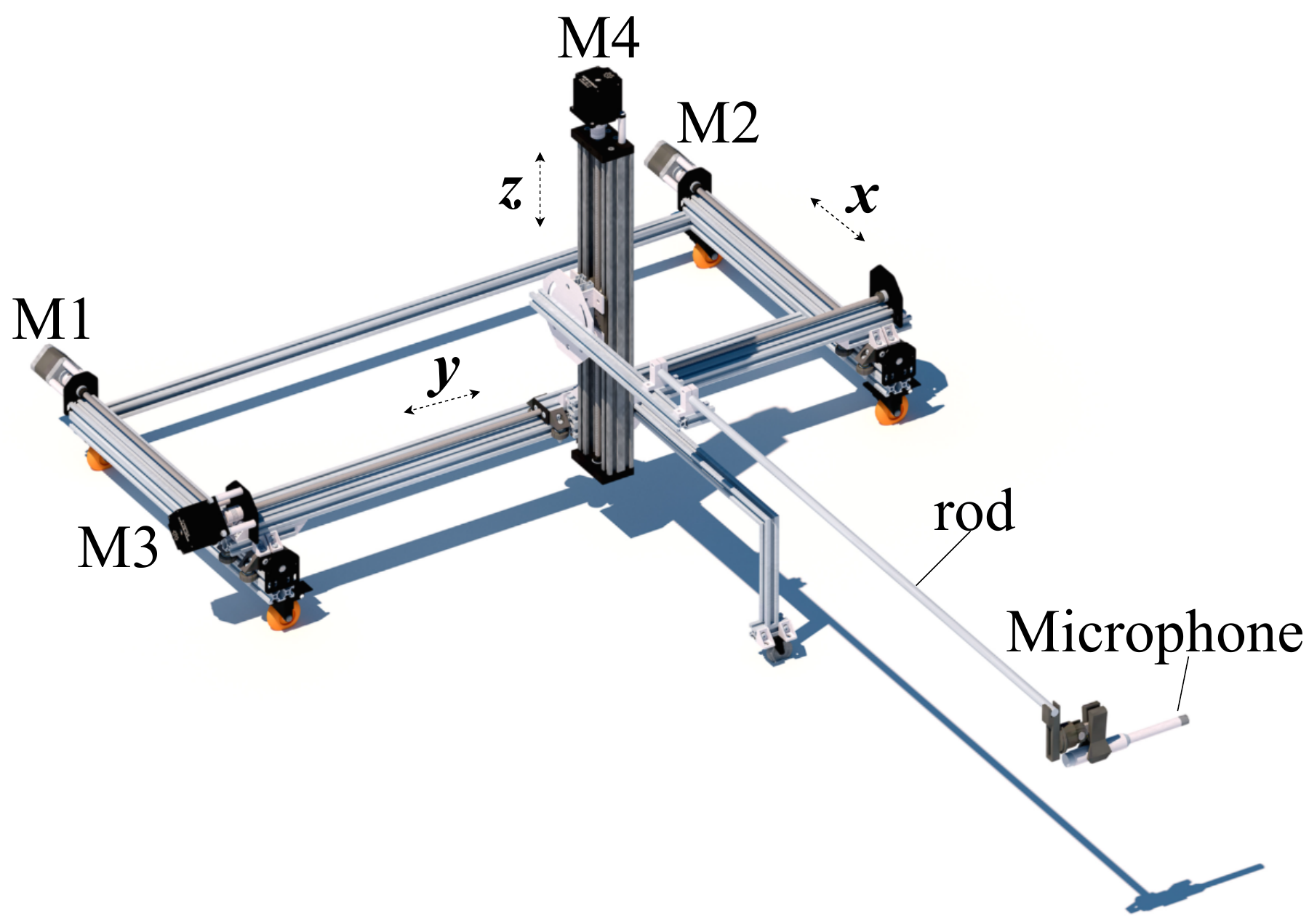}}
\caption{\co (a) The schematics of the FEM/BEM multiphysics simulation used to model finite absorbers and the reconstruction grid of receivers; and (b) schematics of the sound field 3D scanner used in the measurements to position the microphone automatically.}%
\label{fig:scanner_fembem}
\end{figure}

The array measurements reported in this research were performed sequentially. A measurement microphone was moved by a 3D scanner (see \Fig{fig:scanner_fembem-b}) built by the Acoustical Engineering research team from the Federal University of Santa Maria (UFSM, Brazil).  Two identical step motors (\textbf{M1} and \textbf{M2}) operate together to move the base along the $x$ axis. The motor \textbf{M3} moves the base along the $y$ axis, and the motor \textbf{M4} moves it along the $z$ axis. Thus, the rod holding the microphone can be moved with three degrees of freedom. The step motors are controlled by an Arduino, with movement and measurement fully automated using a Python code built by the research team.  

Once the system placed the microphone at the first coordinate of the array, there was a pause of about 8--10~s before measuring the impulse response. This pause allows the vibrations of the rod holding the microphone to cease. After that, a measurement was taken and saved, the microphone was moved to the next coordinate in the array, and the pause and measurement process was repeated until the intended array was fully measured.  

The acoustical signals were acquired with a measurement microphone Behringer ECM~8000, which was connected to a sound card, M-Audio Fast Track Pro (ASIO driver). A full-range loudspeaker mounted on a spherical baffle with a diameter of 9~cm radiated the sound of a logarithmically swept sine. The signal generation, acquisition, and deconvolution were accomplished using the PyTTa package~\cite{fonseca2019pytta,pytta_github}, which is a Python implementation of the ITA~Toolbox~\cite{ITA-Toolbox_2017} for Matlab\textsuperscript{\tiny\textregistered}.

\begin{figure}[ht!]
\centering
\subfloat[]{\label{fig:win_slotted_abs-a}
\includegraphics[width=0.5\linewidth]{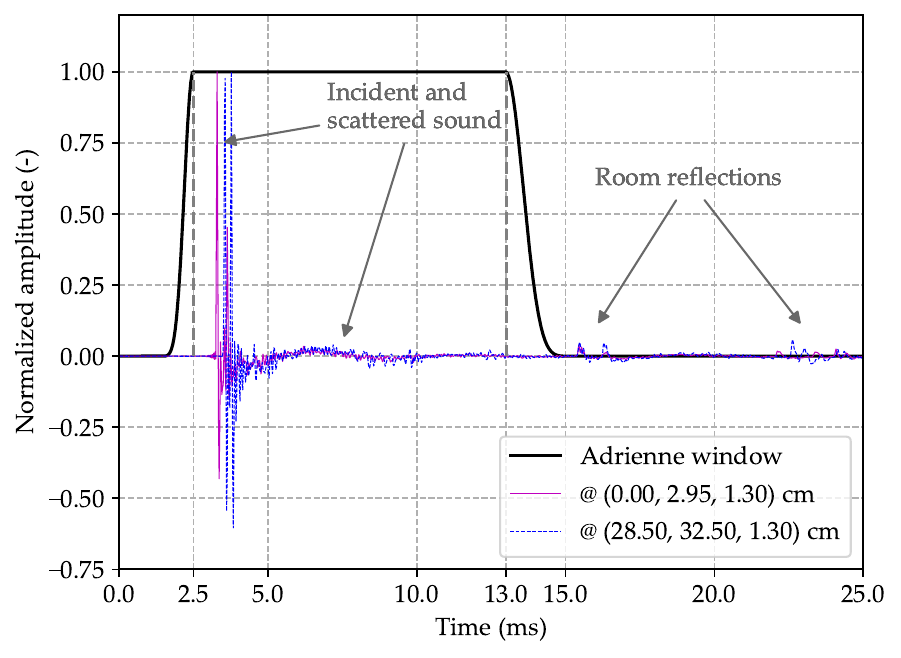}}
\subfloat[]{\label{fig:win_slotted_abs-b}
\includegraphics[width=0.5\linewidth]{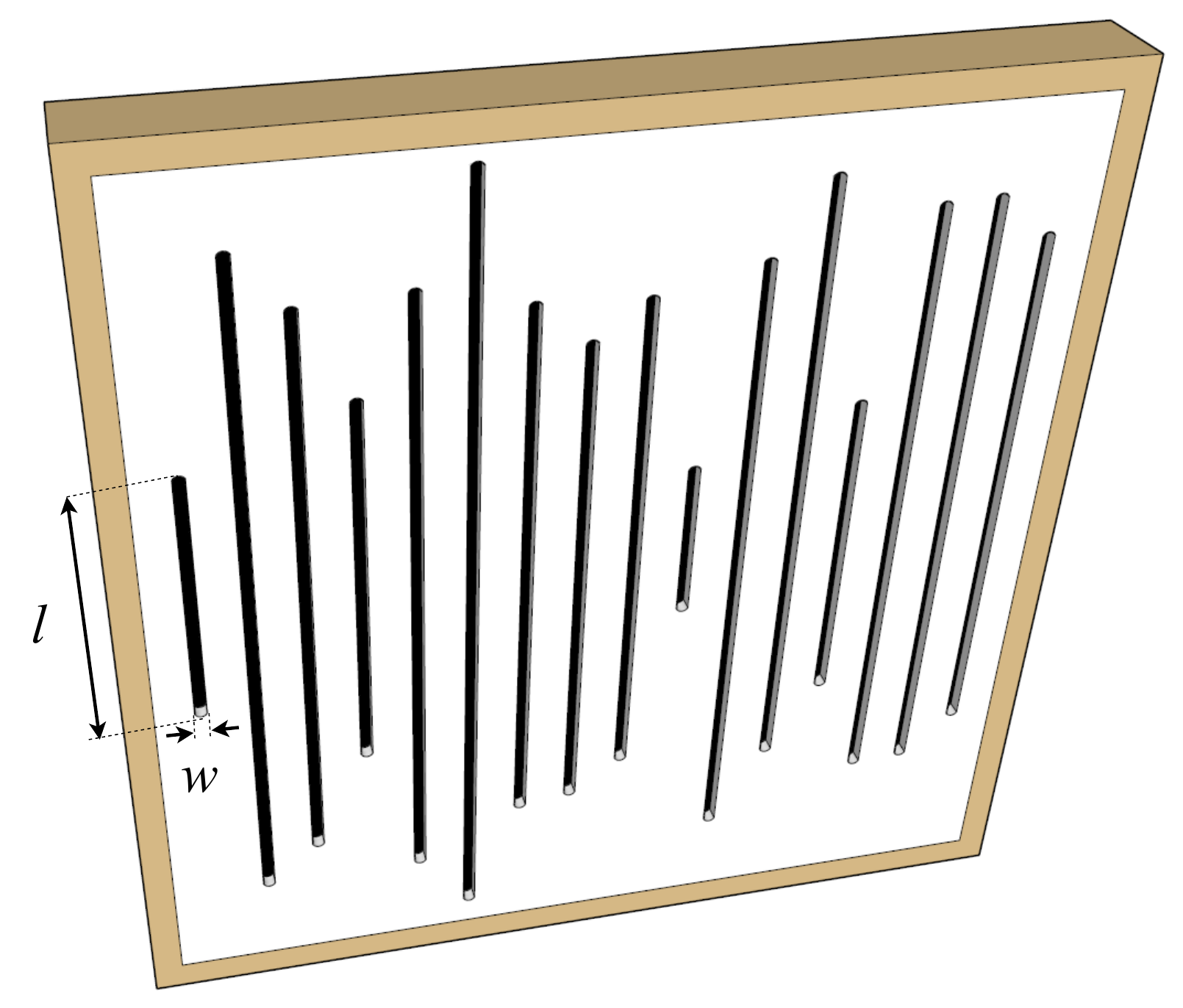}}

\caption{\co (a) The illustration of the windowing process used to separate the incident and scattered sound field components from the room's reflections; and (b) schematics of the Helmholtz resonant absorber with a slit width, $w$, and length, $l$, indicated.}%
\label{fig:win_slotted_abs}
\end{figure}

The measurements were conducted inside a large reverberant room of $207$~m$^3$ of the Acoustical Engineering Laboratory. The sample under measurement was placed at the center of the scanning area, directly on the floor, and in the middle of the room. The sound source was placed at a height of $1.1$~m relative to the floor and, as reported in \Sec{sec:matmet}, the array scans the near field of the absorbers. Since the room is acoustically insulated, it allows for relatively silent measurements. In addition, the reflections in the room emulate a realistic \insitu{} measurement. As the room is relatively large, the distance between the sound source and microphones/sample is significantly shorter than the distance traveled by the reflections on the ceiling and side walls. After the impulse response was acquired, an Adrienne time-window~\cite{garai2000european} was used to remove the undesired parasitic reflections of the room. The principle is to keep the incident sound field and the scattered sound field by the sample while rejecting the undesired room reflections. \Figure{fig:win_slotted_abs-a} shows two measured normalized impulse responses and the Adrienne time-window to illustrate the procedure. The measurements correspond to one receiver near the center of the aperture area of Array~0 and the other receiver on its edge. The spectra of the windowed impulse responses were used as input data for the inverse problem estimation.

Three absorbers were measured in this research: a Polyethylene Terephthalate (PET) fibrous absorber with $60.0 \times 60.0$~cm$^2$, a Melamine foam with $62.5 \times 62.5$~cm$^2$, and a Helmholtz resonant absorber with $65.0 \times 65.0$~cm$^2$. The porous absorbers were placed directly on the floor of the measurement room (considered rigid), not enclosed by a frame during the measurements. The PET wool and the slotted panel were provided \textit{Wave Consultoria}, a room acoustics consulting company in Florianópolis/Brazil. The Melamine and the PET had its flow resistivity ($\sigma$), the porosity  ($\phi$), the tortuosity ($\alpha_{\infty}$), the viscous characteristic length ($\Lambda$), and the thermal characteristic length ($\Lambda^{'}$) measured or estimated. For the PET, the macroscopic parameters were estimated using the inverse method described by Barbosa \myetal~\cite{barbosacob}. For the Melamine foam, the macroscopic parameter measurements were performed at the Acoustical Engineering Laboratory of the Federal University of Santa Maria (UFSM), with the values also reported by Pereira \myetal~\cite{Pereira2021}. The macroscopic parameter values of both samples are given in \Tab{tab:samples}.


\begin{table}[!htb]
	\caption{The thickness and the macroscopic parameters of the porous materials measured in the research.}
	\centering 
	\begin{tabular}{c| c | c |c |c |c |c} 
		\toprule
		 \rowcolor{LightOrange2}  \textbf{Sample} & \textbf{Thickness} ($d$) &  $\sigma$~(Ns/m$^{-4}$) &  $\phi$~(--) & $\alpha_{\infty}$~(--) & $\Lambda$~(\textmu{}m) & $\Lambda^{'}$~(\textmu{}m)\\  
  	 \midrule
		 \textbf{PET} & 5.00~cm & 4683  & 0.90 & 1.00 & 362.1 & 362.2  \\ 
		 \rowcolor[gray]{0.90} \textbf{Melamine} & 3.37~cm & 12200  & 0.98 & 1.01 & 115.0 & 116.0  \\
		\bottomrule
	\end{tabular}
	\label{tab:samples}
\end{table}

The Helmholtz resonant absorber consists of a closed cavity filled with the PET porous absorber ($d = 0.050$~m). On the top of the cavity sits a slotted panel with grooves, as shown in \Fig{fig:win_slotted_abs-b}. The slotted panel has a thickness $t = 0.006$~m, hence the total thickness of the absorber is $0.056$~m. The width of each slot is $w = 0.008$~m, the length, $l$, varies, and the total perforated ratio of the panel is $\Theta = 0.126$. The resonant absorber was fully described and modeled by Barbosa \myetal~\cite{barbosacob}. The authors kindly provided modeling data of the normal incidence absorption coefficient computed with the Transfer Matrix Method (TMM). The equations describing the surface impedance of the Helmholtz absorber were repeated in the \ref{sec:appendixB} for completeness. These data were used for comparison with the \insitu{} measurements. The presentation and discussion of the \insitu{} sound absorption of such a sample is one of this article's contributions since \insitu measurements of such resonant absorbers are difficult to find in the literature.


In all the experiments, the source coordinates were measured relative to the room's floor and left fixed. Therefore, the distance of the source relative to the absorber's center changes with the thickness, $d$, of the absorber and is $\vecb{r}_{\text s} = (0.0, 0.0, 1.1-d)$. Since that was done for all the carried measurements, the models described in \Secs~\ref{sec:matmet:inf} and \ref{sec:matmet:fin} reproduce the experimental set-up.


\subsection{Infinite sample simulation} \label{sec:matmet:inf}

For the first model of the experiment, it is considered that a monopole sound source at the coordinate $\vecb{r}_{\text s} = (x_{\text s}, y_{\text s}, z_{\text s})$ creates the sound field above an infinite absorber ($L_{\text{x}} = L_{\text{y}} = \infty$). Herein, the time dependence is $\e^{\ji \omega t}$, with $\omega$ being the angular frequency and $t$ the independent time variable. This term is omitted throughout the article for the sake of brevity. 

The derivation of the mathematical expression that governs
the sound field created by a monopole near an infinite non-locally reacting absorber was obtained by the Spatial Fourier Transform technique, which, after manipulation, leads to the well-known Sommerfeld Integral~\cite{Wright2005}. Further manipulation of the Sommerfeld Integral allows for numerical evaluation of the sound field at a receiver $\vecb{r}_{\text m} = (x_{\text m}, y_{\text m}, z_{\text m})$, for $z_{\text m} \geq 0$. This model has been addressed in detail in \References~\cite{Brandao2013, Allard1992}, and the resulting sound pressure is
\begin{equation}
\label{eq:sommer}
p(\textbf{r}_\text{m}) = \tilde{s}\left[\frac{\, \e^{-\ji k_{0}\left|\vec{r}_{1}\right|}\,}{\left|\vec{r}_{1}\right|}
-\frac{\, \e^{-\ji k_{0}\left|\vec{r}_{2}\right|}\,}{\left|\vec{r}_{2}\right|} 
+\int_{0}^{\infty} \! \frac{2 \, \rho_{\text{p}} \e^{-\nu_{0 z}\left|z_{\text{s}}+z_{\text{m}}\right|}}{\; \nu_{0 z} \, \rho_{\text{p}}+\nu_{1 z} \, \rho_{0} \tanh \! \left(\nu_{1 z} \, d\right)} \, k \, \text{J}_{0}(k r) \dt k\right].
\end{equation}
%

\Equation{eq:sommer} is valid for a layer of porous material with thickness $d$ over a rigid backing. The complex \wavenumber{}, $k_{\text{p}}$, and the complex density, $\rho_{\text{p}}$, were computed with the Johnson, Champoux, Allard (JCA)~\cite{Johnson1987, Champoux1991} model; $k_0$ and $\rho_0$ are the magnitude of the \wavenumber{} and the density in air, respectively; $\nu_{0 z} = \sqrt{k^{2} - k_{0}^{2}}$, and $\nu_{1 z} = \sqrt{k^{2} - k_{\text{p}}^{2}}$. The term $\tilde{s}$, in kg/s$^{2}$, accounts for the source strength. The distance from the original source to the receiver is $\left|\vec{r}_{1}\right| = \left|\vecb{r}_{\text{s}}- \textbf{r}_{\text{m}}\right|$; the distance from the image source to the receiver is $\left|\vec{r}_{2}\right| = \left|\vecb{r}^{'}_{\text{s}}- \textbf{r}_{\text{m}}\right|$, with $\vecb{r}^{'}_{\text{s}} = (x_{\text s}, y_{\text s}, -z_{\text s})$; $r$ is the horizontal distance given by $r =\sqrt{(x_{\text{s}}-x_{\text{m}})^2+(y_{\text{s}}-y_{\text{m}})^2}$; $\text{J}_{0}$ is the zero-order
Bessel function of the first kind. The particle velocity is computed by Euler's relation, $u_{z}(\textbf{r}_\text{m}) = - \frac{1}{\ji k_0 \rho_0 c_0 } \frac{\partial p(\textbf{r}_\text{m})}{\partial z}$.

\vspace{0.5cm}
The integrals in \Eq{eq:sommer} and the resulting particle velocity still have poles, requiring a special numerical integration routine. The strategy used in this research was to pass a fine grid of possible singularity points to the \texttt{scipy.integrate.quad} (Python) routine \cite{scipy}. If the user supplies enough possible pole positions along the integration path, the numerical calculation of the integral can be done with precision. For $k > k_0$, the integrand decays, and the integral can be truncated at the upper limit of 10 --- a similar approach was used by Brandão \myetal{} \cite{Brandao2013}. It is worth noting that this numerical procedure provides an exact solution to the physical problem at hand, allowing the computation of the sound field at the coordinates of the desired receivers (arrays, center of sample, and the reconstruction grid on the surface). However, numerical computation of this problem is expensive and impractical for inverse problems. Alternative approaches exist, for instance, the one proposed by Eser \myetal{}~\cite{Eser2021}. In contrast, this study proposes a novel inverse problem formulation (\Sec{sec:dcism}) based upon discretizing a more suitable integral equation.


\subsection{Finite sample simulation} \label{sec:matmet:fin}

In order to simulate the diffraction effects on a finite absorber, the Acoustics module of Comsol Multiphysics~6.1 software \cite{comsol} was used. The coupled model built used the Finite Element Method (FEM) interface to account for the sound propagation inside a porous layer of thickness $d$. As in the infinite sample simulation, the JCA~\cite{Johnson1987, Champoux1991} model was used to compute the complex \wavenumber{} $k_{\text{p}}$ and $\rho_{\text{p}}$ of the porous sample. The sound field outside of the sample and its diffraction were computed by the Boundary Element Method (BEM) interface coupled to the FEM interface. The coupling between the two physics interfaces was achieved by the continuity of sound pressure at the top and laterals of the porous sample. The BEM interface also contains an infinite baffle at $z = -d$ and a monopole sound source at $\vecb{r}_{\text{s}}$ (see \Fig{fig:scanner_fembem-a}). The highest simulated frequency was 4~kHz, with at least six elements per wavelength and three elements across the sample thickness. The coordinates of the desired receivers (arrays, center of sample, and the reconstruction grid on the surface) were imported into Comsol as evaluation points. Upon completion of the model execution and the computation of total sound pressures at all evaluation points, the data were subsequently exported and subjected to further processing.


\section{Theory of the inverse problems} \label{sec:theory}

For the formulation of the inverse problems, one must consider the sound field sampled at $\M$ receiver points at $\vecb{r}_{\text m}$, representing the arrays in \Fig{fig:schematics} and \Tab{tab:arrays}. The measured data at each microphone position can be projected into a basis of $\Q$ monopoles with unknown complex amplitudes. Herein, the estimation of these complex amplitudes was achieved by two inverse problems: (i) the Image Source Model (ISM), as proposed by Alkmim \myetal{}~\cite{alkmim2021angle}; and (ii) the Discrete Complex Image Source Model (DCISM), proposed in this work.


\subsection{The Image Source Model (ISM)}\label{sec:ism}

The inherited assumption in this model is used in many \insitu{} impedance estimation methods, with measurements performed either with the two-microphone method~\cite{Li1997}, pressure-velocity probes~\cite{Brandao2015, Brandao2011}, or with a microphone array~\cite{alkmim2021angle}. The model contains the monopole source at $\vecb{r}_{\text s}$, and an image source at $\vecb{r}'_{\text s} = (x_{\text s}, y_{\text s}, -z_{\text s})$. Thus, the number of sources in the model is $\Q = 2$. For a given receiver, the complex amplitude of the sound pressure is given by
\begin{equation}
	\label{eq:ism}
	p(\textbf{r}_\text{m}) = \tilde{s}_{\text{s}} \, G(\vecb{r}_{\text{s}}, \vecb{r}_{\text{m}}) +
        \tilde{s}_{\text{is}} \, G(\vecb{r}'_{\text{s}}, \vecb{r}_{\text{m}}) \,,
\end{equation}
%
%
where $G(\vecb{r}_{\text{s}}, \vecb{r}_{\text{m}}) = \frac{\e^{-\ji k_0 |\textbf{r}_{\text{s}}- \textbf{r}_{\text{m}}|}}{|\textbf{r}_{\text{s}}- \textbf{r}_{\text{m}}|}$ is the free-space Green's Function for the monopole source and the receiver, and $G(\vecb{r}'_{\text{s}}, \vecb{r}_{\text{m}})$ is the corresponding Green's Function for the image source and the receiver. The terms $\tilde{s}_{\text{s}}$ and $\tilde{s}_{\text{is}}$ are the complex amplitudes of the source and image source in kg/s$^2$, respectively. If several receivers measure the sound field, a matrix equation can be written as
\begin{equation}
	\label{eq:ism_matrix}
	\begin{bmatrix} p_1 \\ p_2 \\ \vdots \\ p_{\M} \end{bmatrix} = \begin{bmatrix}  G(\vecb{r}_{\text{s}}, \vecb{r}_{1}) & G(\vecb{r}'_{\text{s}}, \vecb{r}_{1}) \\
G(\vecb{r}_{\text{s}}, \vecb{r}_{2}) & G(\vecb{r}'_{\text{s}}, \vecb{r}_{2}) \\
\vdots & \vdots \\
G(\vecb{r}_{\text{s}}, \vecb{r}_{\M}) & G(\vecb{r}'_{\text{s}}, \vecb{r}_{\M})
\end{bmatrix} \ \begin{bmatrix} \tilde{s}_{\text{s}} \\ \tilde{s}_{\text{is}} \end{bmatrix},
\end{equation}
%
or $\vecb{p} = \vecb{G} \, \vecb{s}$ in matrix form, with $\vecb{p} = \left[p_1, \ p_2, \ \cdots, \ p_{\M}\right]^{\text T} \ \in \ \mathbb{C}^{\M \times 1}$ being the measured vector of complex sound pressures at a given frequency, $\vecb{s} \ \in \ \mathbb{C}^{2 \times 1}$ is the complex amplitude vector (to be determined in an inverse problem --- see \Sec{sec:invest}), and $\vecb{G} \ \in \ \mathbb{C}^{\M \times 2}$ is the kernel matrix containing the Green's Functions of the source and image-source to each receiver.  



\subsection{The Discrete Complex Image Source Method (DCISM)} \label{sec:dcism}

Inherited by the ISM is the assumption that spherical waves reflect on a planar absorber specularly, which is not the case for low frequencies or short source-receiver-sample separations~\cite{Brandao2015,alkmim2021angle}. On the other hand, Di and Gilbert~\cite{Di1993} provided an integral equation with an exact solution for the problem by representing the plane-wave reflection coefficient as the Laplace Transform of an image source distribution along a complex line. Thus, a well-behaved integral was obtained (instead of the hard-to-compute Sommerfeld Integral of \Sec{sec:matmet:inf}). With that formulation, the sound pressure is
\begin{equation}
\label{eq:di1}
	p(\textbf{r}_\text{m})  = 
        \tilde{s}_{\text{s}} \, \frac{\; \e^{-\ji k_0 |\textbf{r}_{\text{s}}- \textbf{r}_{\text{m}}|} \;}{|\textbf{r}_{\text{s}}- \textbf{r}_{\text{m}}|} + 
        \int_{0}^{\infty} \tilde{s}(q) \ G(\vecb{r}_{q}, \vecb{r}_{\text{m}}) \dt q \,,
\end{equation}
with 
\begin{equation}
G(\vecb{r}_{q}, \vecb{r}_{\text{m}}) = \frac{\e^{-\ji\sqrt{r^2+(z_{\text{s}} + z_{\text{m}} -\ji q)^2}}}{\; \sqrt{r^2+(z_{\text{s}} + z_{\text{m}} -\ji q)^2} \;} ,
\end{equation}
%
%
where $r$ is the horizontal distance between the sound source and the $m^{\text{th}}$ receiver, as in \Sec{sec:matmet:inf}. The integral in \Eq{eq:di1} represents a distribution of image sources along a complex line due to the $\ji q$ term in the integrand's Green's Function. This term will render a decaying exponential integrand as $q \rightarrow \infty$. Therefore, the integral is well-behaved and suitable for numerical computation if one knows $s(q)$.

The task proposed in this article is to form an inverse problem suitable to compute a discretized version of the complex amplitudes of the monopoles. The first step towards this goal is modifying \Eq{eq:di1} to deliberately include an image source term associated with the zero at the bottom limit of the integral. That helps to stabilize the inverse problem's solution and is a key aspect in the formulation. The evidence in \Reference~\cite{Di1993} suggests that the forward models for locally and non-locally reacting absorbers have the structure $s(q) = \tilde{s}_{\text{is}} \, \delta(q) + \tilde{s}'(q)$. Therefore, \Eq{eq:di1} can be rewritten as
\begin{equation}
\label{eq:di2}
	p(\textbf{r}_\text{m})  = 
        \tilde{s}_{\text{s}} \, G(\vecb{r}_{\text{s}}, \vecb{r}_{\text{m}}) + 
        \tilde{s}_{\text{is}} \, G(\vecb{r}'_{\text{s}}, \vecb{r}_{\text{m}}) + 
        \int_{0}^{\infty} \tilde{s}'(q) \ G(\vecb{r}_{q}, \vecb{r}_{\text{m}}) \dt q \, \cdot
\end{equation}
%

To form an inverse problem, the integral equation in \Eq{eq:di2} can be discretized by a suitable integration rule. In this research, three discretization schemes are investigated: the mid-point rule (MP), the Gauss-Legendre (GLE) quadrature, and the Gauss-Laguerre (GLA) quadrature. For MP and GLE, a general quadrature rule can be written as
\begin{equation}
\label{eq:quad_mp_leg}
	\int_{0}^{b} f(q) \dt q = \frac{\, b \,}{2} \sum\limits_{i=1}^{N} w_i \, f(\xi_i) \,.
\end{equation}
Usually the weights, $w_i$, and roots, $\xi'_{i}$, are computed for the integration interval from $-1$ to $+1$~\cite{Abramowitz1965}. Since the integrand goes from $0$ to $b$, $\xi_i = \frac{b}{2}\xi'_{i}+\frac{b}{2}$, with $b$ being a suitable truncation value for the integral in \Eq{eq:quad_mp_leg}, and $N$ being the number of integration points in the domain. With the MP rule, the $\xi'_{i}$ mid-points are evenly sampled from $\{-1+2/N\}$ to $\{+1-2/N\}$, and the weights are constant, $w_i = b/N$. For the GLE quadrature, the $w_i$ and $\xi'_{i}$ are computed from Legendre polynomials (see Abramowitz \& Stegun, Ch.~25 \cite{Abramowitz1965}).  In this case, the weights are not constant and the sampling points are not equally spaced along the integration domain.

The Gauss-Laguerre (GLA) quadrature is useful for integrals of the following kind:
\begin{equation}
\label{eq:quad_lag}
	\int_{0}^{\infty} f(q)\e^{-q} \dt q = \sum\limits_{i=1}^{N} w_i \, f(\xi_i) \,,
\end{equation}
%
%
which has the advantage of automatically dealing with the integration interval. However, since the integrand must be of type $f(q) \e^{-q}$, the  integrand in \Eq{eq:di2} must be multiplied by $\e^{+q}$. The increasing exponential is supposed to be compensated by the decreasing term caused by $\ji q$ in the integrand's Green's Function, as long as the order $N$ is kept in check.

Now, one can substitute the integral in \Eq{eq:di2} by the sum in \Eq{eq:quad_mp_leg}, which yields
\begin{equation}
\label{eq:di3}
	p(\textbf{r}_\text{m})  = \tilde{s}_{\text{s}} \, G(\vecb{r}_{\text{s}}, \vecb{r}_{\text{m}}) + \tilde{s}_{\text{is}} \, G(\vecb{r}'_{\text{s}}, \vecb{r}_{\text{m}}) + \frac{b}{2} \sum\limits_{i = 1}^{N} w_i \, \tilde{s}(\xi_{i}) \ G\left(\vecb{r}_{\text{q}}(\xi_i), \vecb{r}_{\text{m}}\right) \,,
\end{equation}
%
with $\vecb{r}_{\text{q}}(\xi_i)$ being the sampled coordinate of the complex image source. If all $\M$ receivers are taken into account, the following system of equations follows
\begin{equation}
	\label{eq:dcism_matrix}
	\begin{bmatrix} p_1 \\ p_2 \\ \vdots \\ p_{\M} \end{bmatrix} = \begin{bmatrix}  G(\vecb{r}_{\text{s}}, \vecb{r}_{1}) & G(\vecb{r}'_{\text{s}}, \vecb{r}_{1}) & \frac{b}{2}  w_1 \, G\left(\vecb{r}_{\text{q}}(\xi_1), \vecb{r}_{1}\right) & \cdots &  \frac{b}{2} w_N G\left(\vecb{r}_{\text{q}}(\xi_N), \vecb{r}_{1} \right) \\
G(\vecb{r}_{\text{s}}, \vecb{r}_{2}) & G(\vecb{r}'_{\text{s}}, \vecb{r}_{2}) & \frac{b}{2} w_1 \, G\left(\vecb{r}_{\text{q}}(\xi_1), \vecb{r}_{2} \right) & \cdots &  \frac{b}{2} w_N \, G\left(\vecb{r}_{\text{q}}(\xi_N), \vecb{r}_{2} \right)\\
\vdots & \vdots & \vdots & \ddots & \vdots\\
G(\vecb{r}_{\text{s}}, \vecb{r}_{\M}) & G(\vecb{r}'_{\text{s}}, \vecb{r}_{\M}) & \frac{b}{2} w_1 \, G\left(\vecb{r}_{\text{q}}(\xi_1), \vecb{r}_{\M} \right) & \cdots &  \frac{b}{2} w_N \, G\left(\vecb{r}_{\text{q}}(\xi_N), \vecb{r}_{\M} \right)
\end{bmatrix} \ \begin{bmatrix} \tilde{s}_{\text{s}} \\ \tilde{s}_{\text{is} } \\ \tilde{s}_{\xi_{1}} \\ \tilde{s}_{\xi_{2}} \\ \vdots \\ \tilde{s}_{\xi_{N}} \end{bmatrix},
\end{equation}
%
%
where  $\textbf{s} \ \in \ \mathbb{C}^{(2+N) \times 1}$ is the vector to be determined by the inverse problem, containing the complex amplitude of the sound source, the image source, and the discrete distribution of the complex image sources (thus $\Q = 2+N$). $\textbf{G}$ is the kernel matrix $\in \ \mathbb{C}^{\M \times (2+N)}$, containing the Green's Functions of the source, the image source, and the complex image sources. The same rationale can be applied to GLA discretization by substituting the integral in \Eq{eq:di2} by the sum in \Eq{eq:quad_lag}. Note that in this case, $b/2 = 1$ in \Eqs{}~\eqref{eq:di3} and \eqref{eq:dcism_matrix}.

\subsection{Inverse problem estimation}\label{sec:invest}

The system of equations $\vecb{p} = \vecb{G} \, \vecb{s}$ can be either overdetermined or underdetermined. In the case of the ISM, $\Q = 2$ and usually $\M \geq 2$ (for array measurements). For the DCISM, $\Q = N+2$ and the number of receivers $\M$ (number of equations) can be either larger or lower than $\Q$ (number of unknowns). However, the condition $\M > \Q$ does not guarantee that $\vecb{p} = \vecb{G} \, \vecb{s}$ is overdetermined, and this can be the case even if $\Q = 2$, as pointed in \Reference~\cite{alkmim2021angle}. The reasons for the system of equations to become ill-posed are noise in the measurement and errors in the model representation. Note also that the matrix $\vecb{G}$ is a function of the array geometry and the Green's Function structure. Thus, symmetrical arrays (\eg Arrays~0 and 1) can create redundancies in matrix $\vecb{G}$. At first sight, this can be regarded as a negative aspect, but as the measured data is somewhat corrupted by noise, the redundancies can carry useful information, leading to less over-fitting of the data.

The least-squares solution to the system of equations is $\vecb{s} = (\vecb{G}^{\text H} \vecb{G})^{-1} \vecb{G}^{\text H} \vecb{p}$, with $\vecb{G}^{\text H}$ being the Hermitian of the matrix $\vecb{G}$. Tests were carried out with such a solution when $\M > \Q$ and the ISM and the DCISM results usually lead to solutions, $\vecb{s}$, with a high $\ell_2$ norm. Eventually, the least-squares solution can lead to pressure and particle velocity reconstructions with a low NMSE, but they seem non-physical. The results are not reported here for the sake of brevity. Instead, a Tikhonov regularized solution is used, given by
\begin{equation}
\label{eq:argmin}
\tilde{\textbf{s}} = \argmin\left(\norm{\vecb{Gs}-\tilde{\vecb{p}}}_{2}^{2} + \lambda^2 \norm{\vecb{s}}_{2}^{2}\right),
\end{equation}		
where $\tilde{s}$ is the estimative of the complex amplitudes of the $\Q$ monopoles and $\lambda>0$ is the regularization parameter, which is estimated herein by the L-curve criterion~\cite{Hansen2010}. Other automatic selection strategies for $\lambda$ exist, such as the generalized cross-validation used in \reference{}~\cite{Richard2019}. Both the solution, $\tilde{s}$ and the regularization parameter, $\lambda$, are found with the aid of the Singular Value Decomposition (SVD) of matrix $\vecb{G}$, which for $\M > \Q$ is given by
\begin{equation}
\label{eq:svdg}
\vecb{G} = \vecb{U} \Sigma \vecb{V}^{H} \,,
\end{equation}
%
%
where $\vecb{U}$ and $\vecb{V}$ are the matrices with the left and the right singular vectors, respectively; and $\Sigma = \text{diag}(\sigma_1, \sigma_2, \cdots, \sigma_{\Q})$ is the diagonal matrix containing the singular values in decreasing magnitude, so that $\sigma_1 \geq \sigma_2 \geq \cdots \geq \sigma_{\Q}$. Therefore, the regularized solution is
\begin{equation}
\label{eq:regsvd}
\tilde{\textbf{s}} = \vecb{V}\left(\Sigma^2 + \lambda^2 \vecb{I}\right)^{-1}  \Sigma \vecb{U}^{H} \vecb{p},
\end{equation}
%
so that $\left(\Sigma^2 + \lambda^2 \vecb{I}\right)^{-1}  \Sigma$ is a diagonal matrix with elements $\frac{\sigma_i}{\sigma^{2}_{i}+\lambda^2}$. Thus, the low-magnitude singular values are filtered out by the regularization parameter. The L-curve parameter tries to find the the value of $\lambda$ that leads to the maximum of the curvature of the function $\text{log}(\norm{\tilde{s}}_2) \times \text{log}(\norm{\vecb{G} \tilde{s}- \vecb{p}}_2)$ (solution $\ell_2$ norm vs. residual $\ell_2$ norm). Instead of doing this by trial and error,  the SVD of matrix $\vecb{G}$ and the measured data $\vecb{p}$ are used for the computation of a suitable value of $\lambda$ (see Hansen, Chs.~4--5 \cite{Hansen2010}).

\subsection{Sound field reconstruction}\label{sec:recon}

Once $\tilde{\textbf{s}}$ is estimated, one can reconstruct the sound field elsewhere in the vicinity of the absorbing sample. The reconstructed sound pressure is given by~\cite{Nolan2019}
\begin{equation}
\label{eq:pHxre}
\tilde{\vecb{p}}_{\text{re}} = \vecb{G}_{\text{re}} \ \tilde{\textbf{s}} \,,
\end{equation}
%
%
where $\tilde{\vecb{p}}_{\text{re}} \ \in \mathbb{C}^{K}$ is the reconstructed sound pressure vector estimated at a set of $K$ positions and $\vecb{G}_{\text{re}} \ \in \mathbb{C}^{K \times \mathcal{L}}$ is the reconstruction matrix containing the Green's Functions evaluated at the reconstruction points. The particle velocity at $z$ direction can be calculated from Euler’s equation of motion as
\begin{equation}
\label{eq:uHxre}
\tilde{\vecb{u}}_{\text{z-re}} = \frac{-1}{\; \ji \, k_0 \, \rho_0 \, c_0 \;} \ \frac{\partial\vecb{G}_{\text{re}}}{\partial z} \ \tilde{\textbf{s}} \,,
\end{equation}
%
%
where $\frac{\partial\vecb{G}_{\text{re}}}{\partial z} \ \in \mathbb{C}^{K \times \mathcal{L}}$ contains the partial derivative of the reconstruction matrix with respect to $z$. For \insitu{} measurements, it is interesting to reconstruct the sound pressure and the $z$ component of the particle velocity at a grid of points at the surface of the sample. Thus, the reconstructed surface impedance can be estimated as the spatial average of several reconstructed surface impedances. Typically, the grid is small enough to ensure that the spatial variation of $Z\txu{s}$ in the frequency range of interest is not too significant. The grid used in this research, defined in \Sec{sec:matmet}, is the same one used in \References~\cite{Brandao2022, Richard2017}.
\section{Results and discussion} \label{sec:results}

This section presents the results and analysis of the \insitu{} experiments and the simulations. 
The simulations were conducted in the frequency domain, utilizing discrete central frequencies corresponding to the 1/6$^{\text{th}}$ octave bands. The experimental results are shown with a 5~Hz frequency resolution and the frequency range is limited from 100~Hz to 4~kHz. The ISM (\Sec{sec:ism}) and the DCISM (\Sec{sec:dcism}) are compared for Arrays~0--3 (see \Tab{tab:arrays} and \Fig{fig:schematics}). For the DCISM, the discretization schemes are also investigated (Gauss-Legendre - GLE, Gauss-Laguerre - GLA, and Mid-Point - MP samplings); the hyper-parameters are $b = 30$ (upper limit of integration) and $N=25$ (number of Gauss points) unless stated otherwise. For the simulations, the NMSE in \Eq{eq:nmse} is used to assess the errors in the reconstructed pressure and the $\hat{z}$ component of particle velocity. The absorption coefficient is also evaluated for the \insitu{} experiments and compared to the absorption coefficients obtained from \planewave{} incidence (theoretical, \Eqs{}~\eqref{eq:alpha} and \eqref{eq:zs}) and spherical wave incidence (simulations with $Z\txu{s}$ obtained at the sample's surface center). 
%

\subsection{Infinite sample simulation} \label{sec:results_inf}

\begin{figure}[ht!]
\centering
\subfloat[]{\label{fig:pet_inf_nmse-a}
\includegraphics[width=0.49\linewidth]{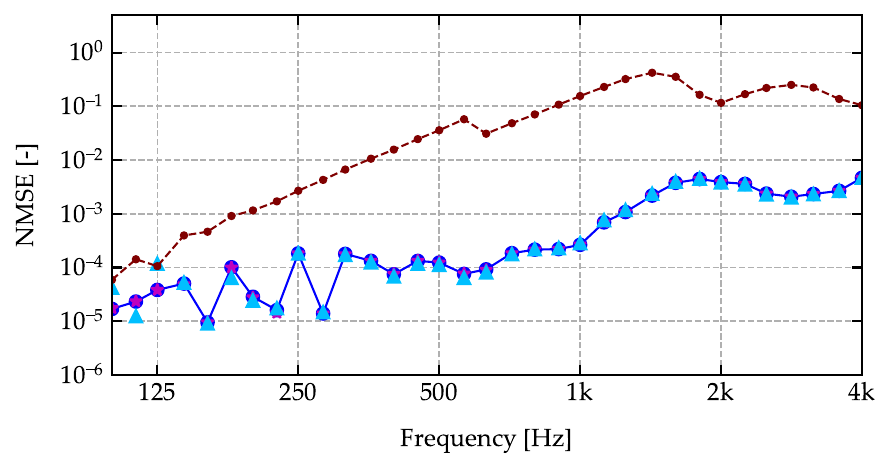}}
\subfloat[]{\label{fig:pet_inf_nmse-b}
\includegraphics[width=0.49\linewidth]{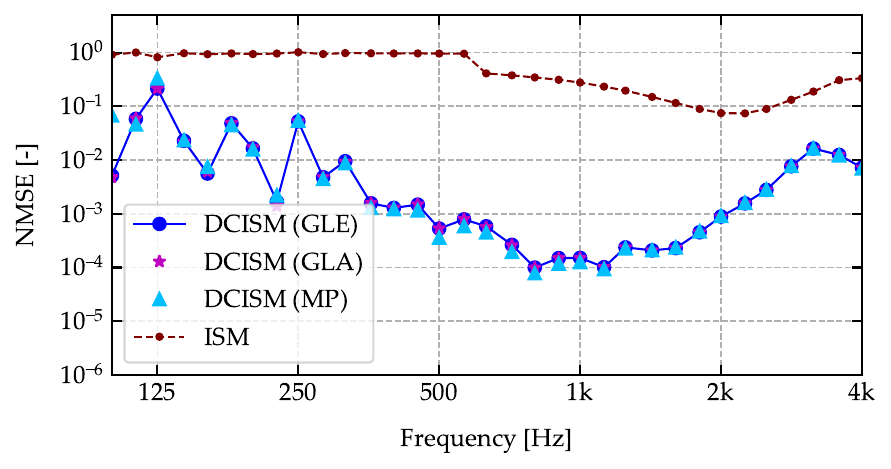}}
\caption{\co The NMSE of the reconstructions for the DCISM (GLE, GLA, and MP discretizations) and the ISM for the simulation of the measurement above an infinite PET porous absorber: (a) NMSE for the sound pressure reconstruction; and (b) NMSE for the $\hat{z}$ component of the particle velocity reconstruction.}%
\label{fig:pet_inf_nmse}
\end{figure}

The simulation of the experiment of the infinite and non-locally reacting sample was described in \Sec{sec:matmet:inf}. \Figure{fig:pet_inf_nmse} presents the NMSE vs. frequency obtained for the PET porous absorber with measurement performed with Array~0. The discretization schemes of the DCISM are compared to the ISM. For the sound pressure (\Fig{fig:pet_inf_nmse-a}), the NMSE of all DCISM discretization schemes is below $0.01$ for all frequencies. At very low frequencies, the NMSE of the DCISM with MP sampling can be slightly higher, and there are no significant differences between the DCISM with GLE and the GLA sampling schemes. The NMSE obtained from the ISM method is significantly higher than for all DCISM cases, reaching a maximum of $\approx 0.42$ and being higher than $0.1$ above 900~Hz. These high NMSE values are attributed to the simplicity of the underlying assumption of the ISM, which considers only the source and an image source.

\vspace{0.5cm}
The NMSE for particle velocity, shown in \Fig{fig:pet_inf_nmse-b}, is generally higher than for sound pressure. For the ISM, it is above $0.3$ below 1~kHz, leading to an inaccurate particle velocity prediction. It is significantly lower for all the DCISM sampling schemes, below $0.01$ between 225~Hz and 2.5~kHz. However, the NMSE of particle velocity increases as the frequency decreases, which happens because the actual value of the particle velocity at the surface tends to be small for porous absorbers in the low-frequency range as a porous absorber tends to be more reflective. 
Additionally, deviations arise when reconstructing the particle velocity from pressure data~\citep{Jacobsen2005}.

\begin{figure}[ht!]
\centering
\subfloat[]{\label{fig:pet_inf_alpha_s-a}
\includegraphics[width=0.49\linewidth]{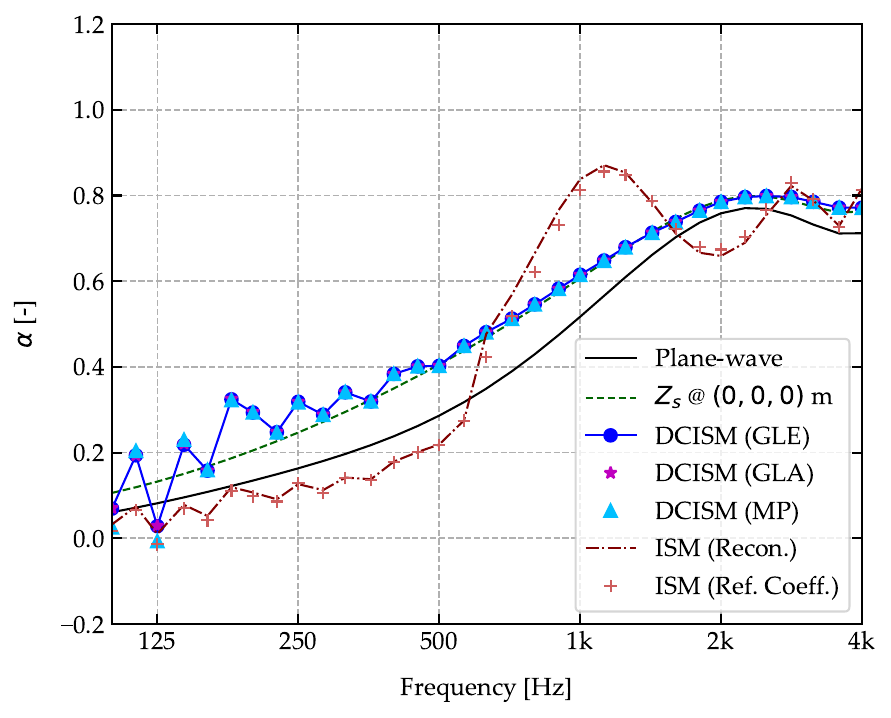}}
\subfloat[]{\label{fig:pet_inf_alpha_s-b}
\includegraphics[width=0.49\linewidth]{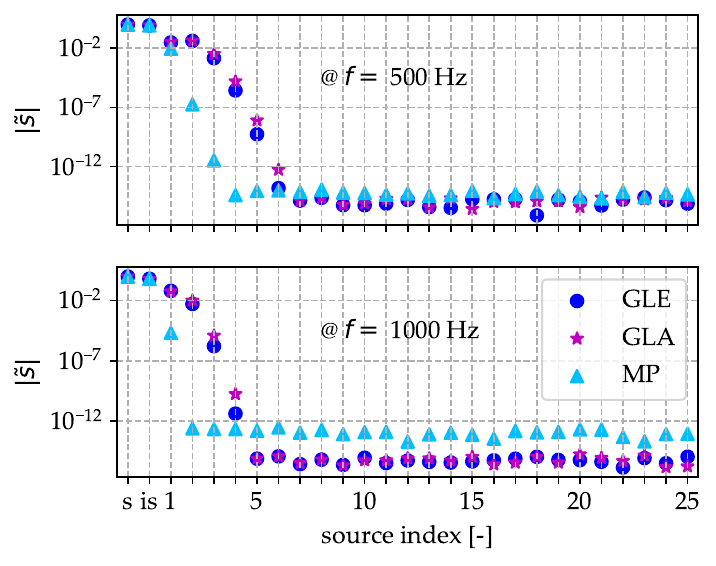}}
\caption{\co Simulation of the measurement of an infinite PET porous absorber: (a) sound absorption coefficient obtained from Array~0 with the DCISM (GLE, GLA, and MP discretizations) and with the ISM (reconstruction and reflection coefficient) --- for reference, the \planewave{} absorption coefficient and the one from spherical wave incidence; and (b) magnitude, $|\tilde{s}|$, of the sources for the DCISM (GLE, GLA, and MP discretizations) for 500~Hz (top) and 1~kHz (bottom).}%
\label{fig:pet_inf_alpha_s}
\end{figure}

\Figure{fig:pet_inf_alpha_s-a} displays the predicted absorption coefficient of the PET sample for several cases. The \planewave{} absorption coefficient is underestimated relative to the one obtained from the surface impedance, $Z_{\text s}$ measured at $(0, 0, 0)$. The reason for this is the non-local reaction behavior of the PET sample, with a low flow resistivity of $\sigma = 4683$~Ns/m$^{4}$. The absorption coefficient obtained with the GLE, GLA, and MP discretization schemes of DCISM matches the one obtained from $Z_{\text s}$ measured at $(0, 0 ,0)$ above 225~Hz. The agreement indicates that the DCISM can predict the absorption caused by the spherical wave sound field.

For completeness, the absorption coefficient of the ISM prediction is computed in two ways for this case. First by reconstruction of the sound field (Recon.), as with the DCISM, and also from the reflection coefficient, as done in \Reference~\cite{alkmim2021angle} with $\alpha = 1 - |\tilde{s}_{\text{is}}/\tilde{s}_{\text{s}}|^{2}$ (Ref. Coeff.). Note in \Fig{fig:pet_inf_alpha_s-a} that the absorption coefficient predicted with the ISM by both methods is very similar but does not match any reference curve, being underestimated at lower frequencies. Such errors tend to decrease if the source-sample distance increases, as discussed in \reference{}~\cite{alkmim2021angle}. It is also noteworthy that if the ISM curves were presented without a reference, they could be considered plausible absorption coefficients for a porous sample. Nonetheless, they are not accurate, which is attributed to the simplicity of the underlying assumption built into the model.

\Figure{fig:pet_inf_alpha_s-b} shows the magnitude of the source and image sources for the DCISM (GLE, GLA, and MP) at 500~Hz (top) and 1~kHz (bottom). The estimations made by the GLE and GLA agree relatively well. Note also that for such discretization schemes, there are at least six active sources ($|\tilde{s}|> 10^{-7}$ --- source, image source, and 4 complex sources). That indicates the importance of including a distribution of monopoles in the sound field model and further highlights the simplicity of the ISM. For the MP discretization scheme, the estimated source magnitudes decay more abruptly with source order, indicating less active sources. A word of caution about the GLA discretization is necessary. One should note that the upper limit of integration is chosen automatically and that the integrand is pre-multiplied by $\e^{+q}$. Thus, as the number of integration points increases, the upper limit of integration also increases and, consequently, the maximum value of $\e^{+q}$. Suppose the integration order is set too high. In that case, the integrand's decaying term can no longer compensate for the increase in $\e^{+q}$, and the integrand diverges, leading to a non-convergent inverse problem. For such reasons, only the GLE discretization scheme will be used for the DCISM from this point onward.


\begin{figure}[ht!]
\centering
\subfloat[]{\label{fig:melamine_inf_nmse-a}
\includegraphics[width=0.49\linewidth]{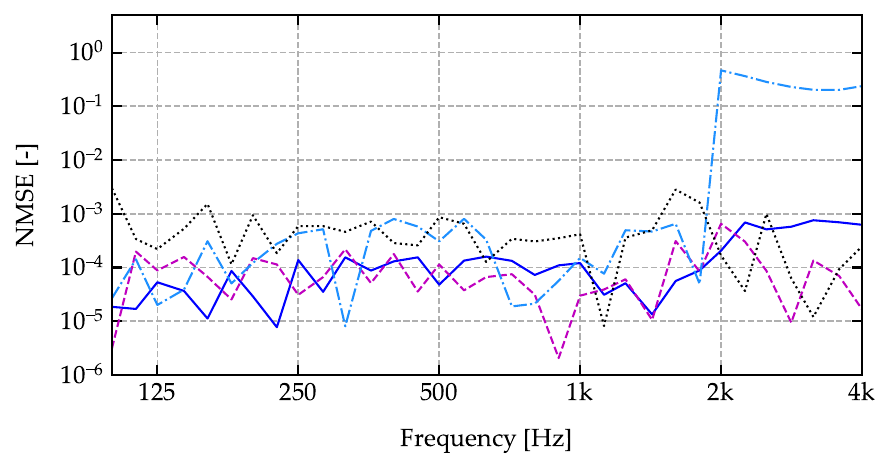}}
\subfloat[]{\label{fig:melamine_inf_nmse-b}
\includegraphics[width=0.49\linewidth]{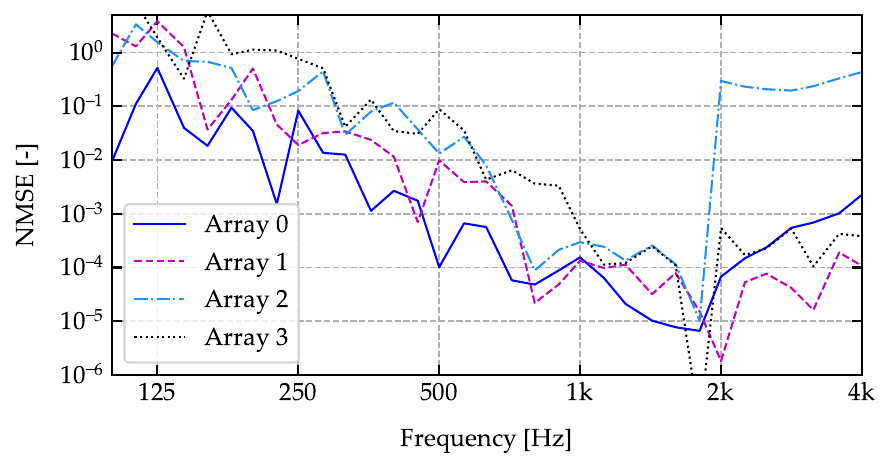}}
\caption{\co The NMSE of the reconstructions for Arrays~0--3 for the DCISM (GLE discretization) for the simulation of the measurement above an infinite Melamine porous absorber: (a) NMSE for the sound pressure reconstruction; and (b) NMSE for the $\hat{z}$ component of the particle velocity reconstruction.}%
\label{fig:melamine_inf_nmse}
\end{figure}

\Figure{fig:melamine_inf_nmse} presents the NMSE vs. frequency obtained for the infinite Melamine porous absorber with measurements performed with Arrays~0--3 and the DCISM. For the sound pressure (\Fig{fig:melamine_inf_nmse-a}), the NMSE for Arrays~0, 1, and 3 are below $0.001$ for almost all frequencies, increasing slightly above 1.5~kHz. For the particle velocity (\Fig{fig:melamine_inf_nmse-b}), the NMSE shows higher values at lower frequencies, also attributed to the low value of the actual particle velocity. For both quantities, the lower NMSE is generally found for Arrays~0 and 1, which correspond to the larger aperture areas and number of sensors, indicating that the more data gathered, the better (at least for the infinite sample). For Array~2, there is a sudden increase in the NMSE above 2~kHz, a feature also found for the simulation of the finite sample (see \Sec{sec:results_fin}) and in the experimental absorption coefficient data (see \Sec{sec:results_meas}). The reason for that remains unknown, but the hypothesis is that it has to do with the array geometry, which impacts the conditioning of the matrix $\vecb{G}$. That indicates the need for more research on the influence of the array geometry on measurement accuracy.

\begin{figure}[ht!]
\begin{center}
\includegraphics[width=0.6\columnwidth]{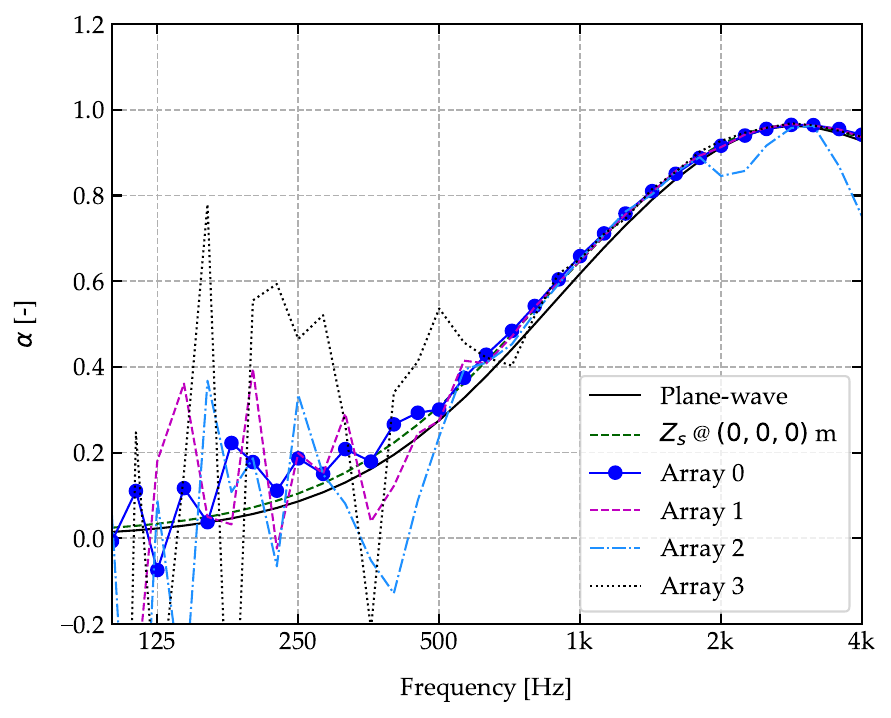}
\caption{\label{fig:melamine_inf_alpha}{\co The sound absorption coefficient for the simulation of the measurement of an infinite Melamine porous absorber obtained from the DCISM (GLE discretization) with Arrays~0--3; for reference, the \planewave{} absorption coefficient and the one from spherical wave incidence.}}
\end{center}
\end{figure}

\Figure{fig:melamine_inf_alpha} displays the absorption coefficient of the Melamine foam, obtained with the DCISM for Arrays~0--3. The estimated data is compared to the absorption coefficient for \planewave{} incidence and spherical wave incidence measured at the sample's center. For the Melamine, with a higher flow resistivity than the PET, the difference between plane and spherical wave incidence is less significant for the investigated source distance ($\approx 1.10$~m). For Array~0, the DCISM estimation agrees well with the spherical wave incidence above 225~Hz. For the other arrays, the agreement is better above $\approx 500$~Hz, with poorer agreement at lower frequencies. Furthermore, the smaller the number of sensors in the array, the more the sound absorption estimation seems to be corrupted by noise. Note also the disagreement with the reference data for Array~2 above 2~kHz, which is linked to the sudden increase in the NMSE for such array (see \Fig{fig:melamine_inf_nmse}).

\subsection{Finite sample simulation} \label{sec:results_fin}

\begin{figure}[ht!]
\centering
\subfloat[]{\label{fig:melamine_fin_nmse-a}
\includegraphics[width=0.49\linewidth]{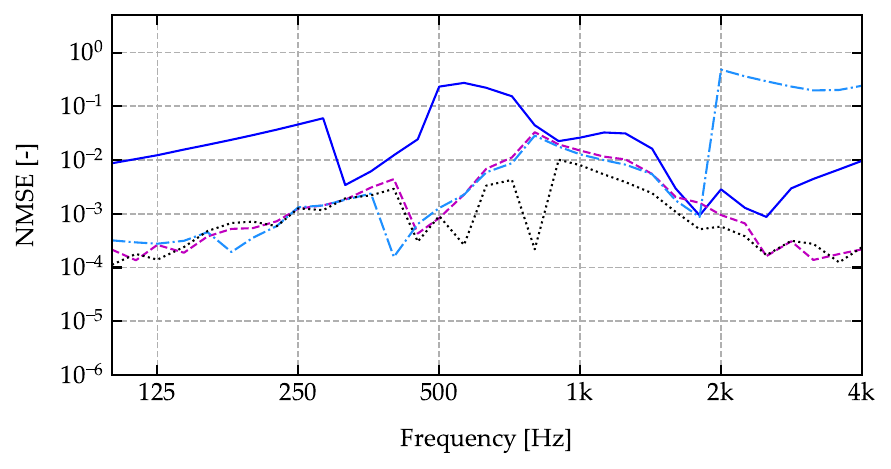}}
\subfloat[]{\label{fig:melamine_fin_nmse-b}
\includegraphics[width=0.49\linewidth]{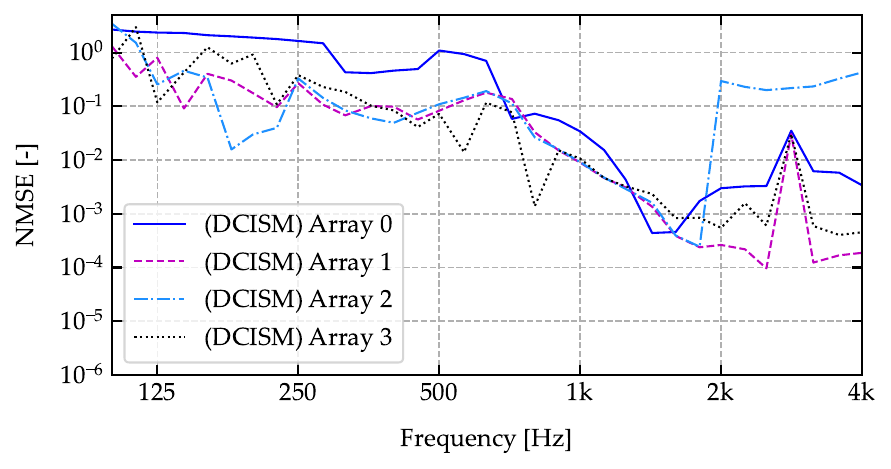}}\\
\subfloat[]{\label{fig:melamine_fin_nmse-c}
\includegraphics[width=0.49\linewidth]{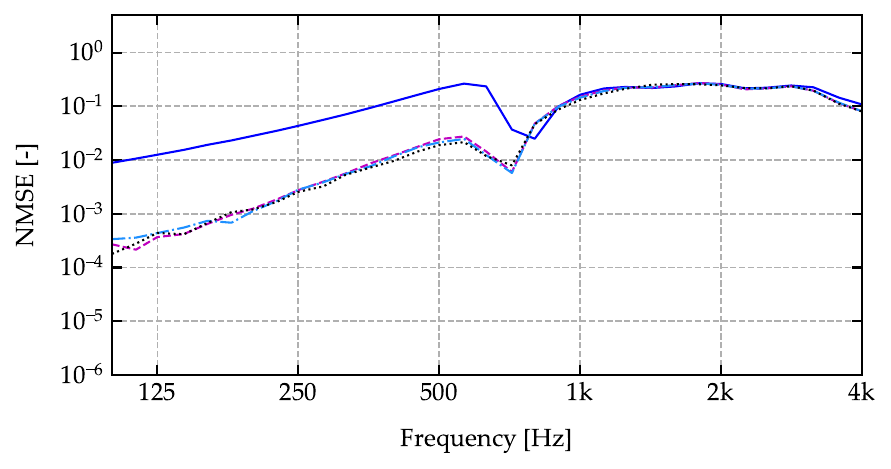}}
\subfloat[]{\label{fig:melamine_fin_nmse-d}
\includegraphics[width=0.49\linewidth]{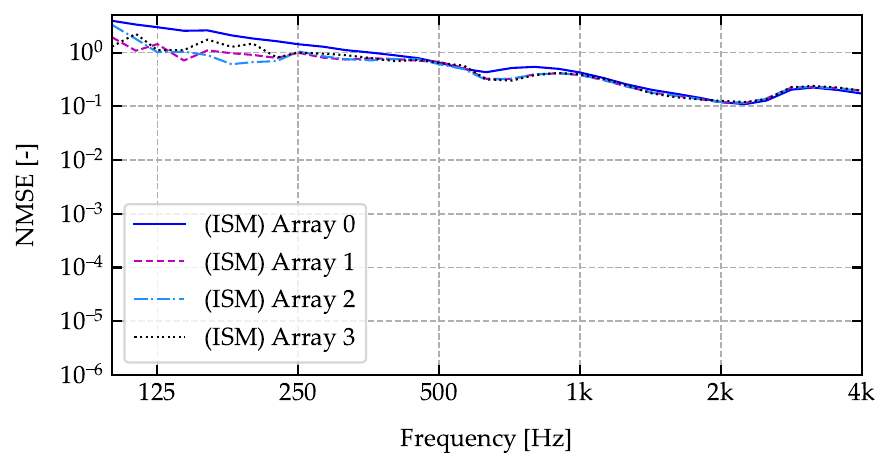}}
\caption{\co NMSE of the reconstructions for Arrays~0--3 for the simulation of the measurement above a finite Melamine porous absorber: (a) and (b) NMSE for the sound pressure and the $\hat{z}$ component of the particle velocity reconstructions for the  DCISM (GLE discretization), respectively; and (c) and (d) NMSE for the sound pressure and the $\hat{z}$ component of the particle velocity reconstructions for the  ISM, respectively.}%
\label{fig:melamine_fin_nmse}
\end{figure}

The simulation of the experiment of the finite and non-locally reacting sample was described in \Sec{sec:matmet:fin}. \Figure{fig:melamine_fin_nmse} presents the NMSE vs. frequency obtained for the finite Melamine porous absorber for measurements performed with Arrays~0--3, with the DCISM (\Figs{} \ref{fig:melamine_fin_nmse-a} and \ref{fig:melamine_fin_nmse-b}), and with the ISM (\Figs{} \ref{fig:melamine_fin_nmse-c} and \ref{fig:melamine_fin_nmse-d}). Comparing the DCISM and the ISM, it is noteworthy that the NMSE of both pressure and particle velocity is lower for the DCISM in all array configurations, even more above 1~kHz. Therefore, the DCISM is more accurate in reconstructing the sound field on the surface of this finite absorber. With the simulation of the experiment as it is, Array~0 is the one with the worst performance, with errors below $0.1$, only above 630~Hz for the DCISM, which is attributed to the size of Array~0 ($0.57 \times 0.65$~m$^2$), relative to the size of the sample ($0.625 \times 0.625$~m$^2$). Thus, Array~0 has a significant part of its aperture near the edges of the sample and captures the edge diffraction effects that the underlying assumptions in the inverse problems cannot handle. Arrays~1--3 have lower NMSE than Array~0 (see \Tab{tab:arrays} for their size), with Array~3 performing slightly worse due to the limited number of microphones and Array~2 exhibiting the sudden increase in the NMSE near 2~kHz (a feature also observed in the simulations of the experiment near the infinite absorbers --- see \Fig{fig:melamine_inf_nmse}).

\begin{figure}[ht!]
\centering
\subfloat[]{\label{fig:melamine_fin_alpha-a}
\includegraphics[width=0.49\linewidth]{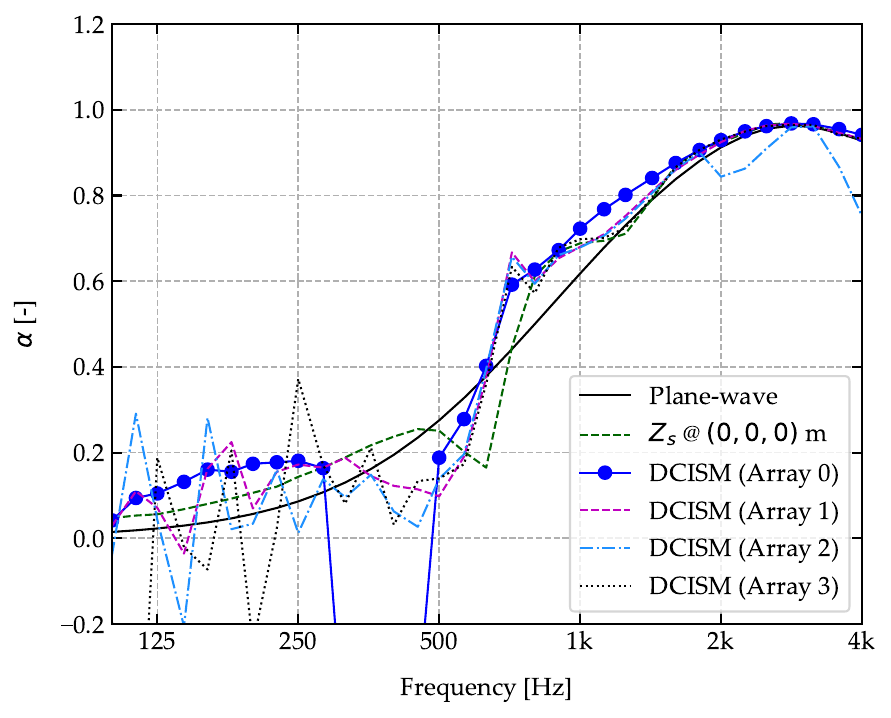}}
\subfloat[]{\label{fig:melamine_fin_alpha-b}
\includegraphics[width=0.49\linewidth]{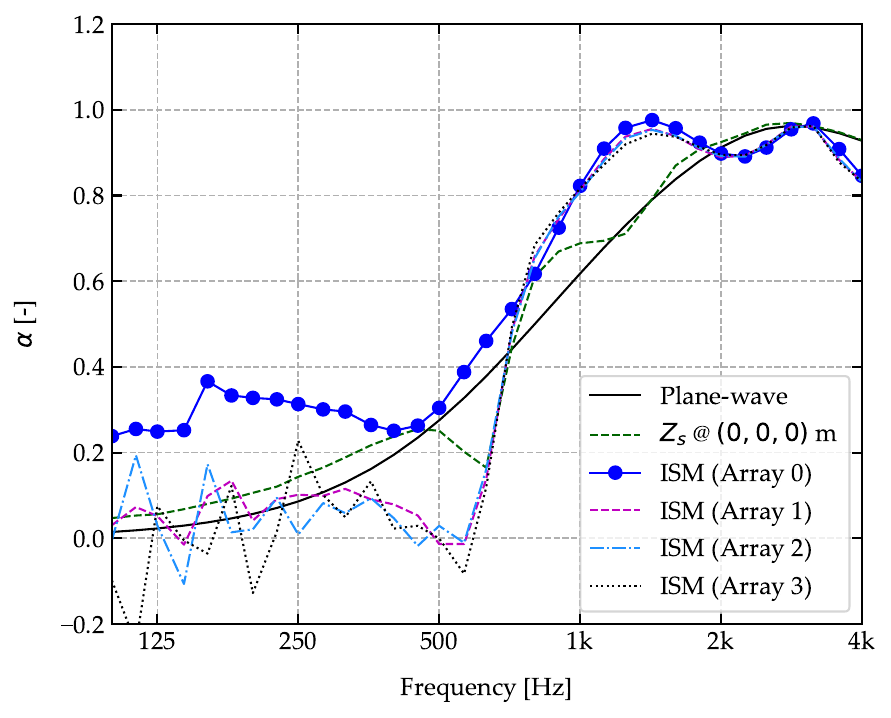}}
\caption{\co Sound absorption coefficient for the simulation of the measurement of an infinite Melamine porous absorber obtained with Arrays~0--3 --- for reference, the \planewave{} absorption coefficient and the one from spherical wave incidence; (a) processed with the DCISM (GLE discretization); and (b) processed with the ISM.}%
\label{fig:melamine_fin_alpha}
\end{figure}

\Figure{fig:melamine_fin_alpha} shows the absorption coefficient of the finite Melamine foam, obtained with the DCISM and ISM for Arrays~0--3. The estimated data is compared to the absorption coefficient for \planewave{} incidence (theoretical infinite sample) and for spherical wave incidence measured at the finite sample's center. Here, the sample size effect is evident through the oscillations of the green curve ($Z_{\text s}$ @ $(0, 0, 0)$). This effect was described in \References~\cite{Brandao2012, Hirosawa2009}. It should be observed that for DCISM (as shown in \Fig{fig:melamine_fin_alpha-a}), the absorption coefficient estimated using Arrays~1--3 aligns with the value derived from $Z_{\text s}$ at the center of the sample for frequencies above 500~Hz --- this includes the oscillations attributable to the effect of sample size.
For Array~0, the agreement is poorer, especially below 500~Hz. Note also the disagreement with the reference data for Array~2 above 2~kHz, which is linked to the sudden increase in the NMSE for such an array. The same discussion applies to the ISM, but note that the disagreement with the reference curves is more exaggerated, caused by the underlying model's relative simplicity. Similar trends were observed for the PET sample but are not reported here for brevity.

\vspace{0.5cm}
The fact that the DCISM reconstructions can match the data from $Z_{\text s}$ at the sample's center is worth noting. Since the actual value of $Z_{\text s}$ is different for the infinite and the finite samples, one can conclude that the method can reconstruct the particular sound field quantity with adequate precision. Note also that the surface impedance can vary in space for a finite and non-locally reacting sample, which is not the case for a locally reacting absorber, as investigated by Brandão \myetal{} \cite{Brandao2022, Brandao2012}.

Here, a brief discussion and anticipation of the experimental results are necessary. The simulated sample size effects above porous absorbers are somewhat exaggerated relative to the experimental results in \Sec{sec:results_meas}. At this point in the research, it is not fully understood to which extent the BEM/FEM multiphysics simulation deviates from the experimental conditions. For instance, in the simulations, the sound source is omnidirectional, whereas some directivity exists in a realistic loudspeaker sound source. Additionally, in the simulations, the sample has sharp edges and is considered isotropic, which might not correspond to reality. Time-windowing also plays a role in the post-processing of the experimental data, which requires better understanding. In the literature, it has also been reported that the lateral boundary conditions of the sample play a role in the edge diffraction effect observed in the \insitu{} measurement of the sound absorption coefficient~\cite{Brandao2011a}.    

Furthermore, the effects of adding Gaussian noise to the simulations also seem exaggerated relative to the experiments (which shows less variation across the arrays). That is a positive aspect since experiments are expected to be less noise-sensitive than what was simulated. The fact is that more research is necessary both on the simulation and on the experimental fronts, particularly for finite absorbers, which are tasks to pursue in the near future. 

\subsection{Measurements} \label{sec:results_meas}

\begin{figure}[ht!]
\centering
\subfloat[]{\label{fig:petmel_meas_dcismvsism-a}
\includegraphics[width=0.49\linewidth]{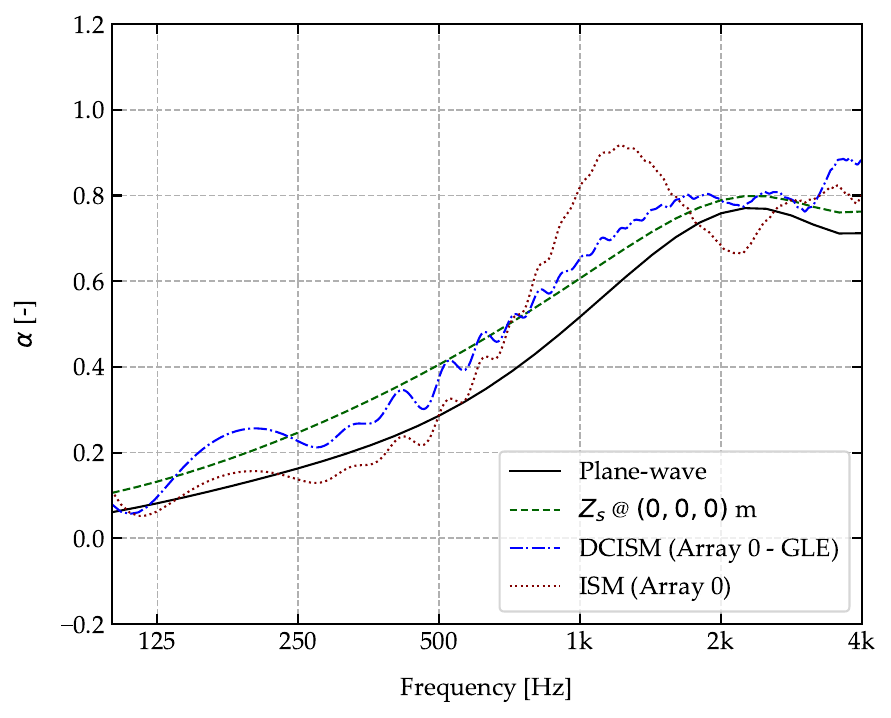}}
\subfloat[]{\label{fig:petmel_meas_dcismvsism-b}
\includegraphics[width=0.49\linewidth]{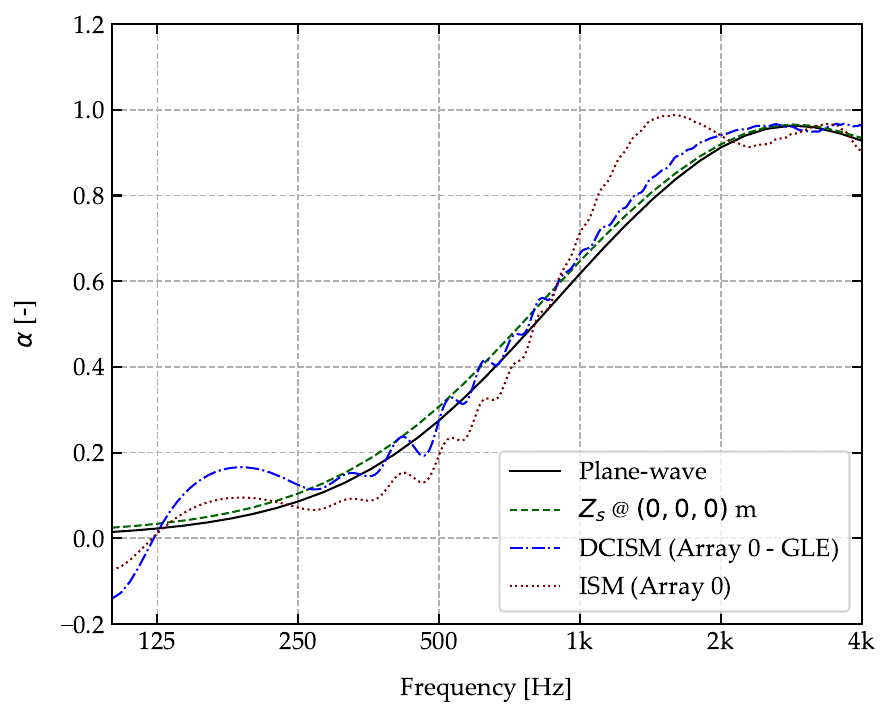}}
\caption{\co Sound absorption coefficient for the measurement of two finite porous absorbers obtained from Array~0 with the DCISM (GLE discretization and the ISM) --- for reference, the \planewave{} absorption coefficient and the one from spherical wave incidence on the infinite sample; (a) PET porous absorber; and (b) Melamine porous absorber.}%
\label{fig:petmel_meas_dcismvsism}
\end{figure}

This section presents the results of the experiments described in \Sec{sec:matmet:exp}. \Figure{fig:petmel_meas_dcismvsism} displays the predicted absorption coefficient for the PET and the Melamine porous absorber. The theoretical \planewave{} absorption coefficient (black curve) and the one estimated from the surface impedance, $Z_{\text s}$, measured at $(0, 0, 0)$ at the infinite sample (green curve - spherical wave incidence) are plotted for comparison. Recall that for the PET, the flow resistivity is $\sigma = 4683$~Ns/m$^{4}$ and for the Melamine $\sigma = 12200$~Ns/m$^{4}$. Therefore, the differences between the \planewave{} and spherical wave absorption coefficients are more significant for the PET sample. The experimental curves were obtained from data collected by Array~0 and processed with the DCISM (GLE discretization, blue curve) and the ISM (brown curve). The absorption coefficient obtained with the DCISM matches the one from $Z_{\text s}$ measured at $(0, 0, 0)$ above 300~Hz, which is more overestimated relative to the \planewave{} absorption coefficient for the PET sample, with a lower flow resistivity (\Fig{fig:petmel_meas_dcismvsism-a}). Thus, the DCISM can predict the absorption caused by the spherical wave incidence with accuracy. The agreement for the Melamine (\Fig{fig:petmel_meas_dcismvsism-b}) is also satisfactory, but the difference between plane and spherical wave incidence is less critical. For the PET, the absorption coefficient estimated with the ISM has a similar trend to the one in \Fig{fig:pet_inf_alpha_s-a}, confirming that its simplicity leads to an absorption coefficient that is plausible but that does not match any theoretical value. Such an effect seems less critical for the Melamine, but it still exists.

The sample size effect is less prominent for both porous absorbers than in the simulations shown in \Sec{sec:results_fin}. Note that the experimental curves obtained from the DCISM tend towards the infinite sample simulation in both \Figs{}~\ref{fig:petmel_meas_dcismvsism-a} and \ref{fig:petmel_meas_dcismvsism-b} (for spherical wave incidence). As discussed before, more research is necessary on this topic, but the fact that the DCISM can render a reliable absorption coefficient on such experimental data is regarded as a positive aspect of the proposed method.

\begin{figure}[ht!]
\centering
\subfloat[]{\label{fig:petmel_meas_arraycomp-a}
\includegraphics[width=0.49\linewidth]{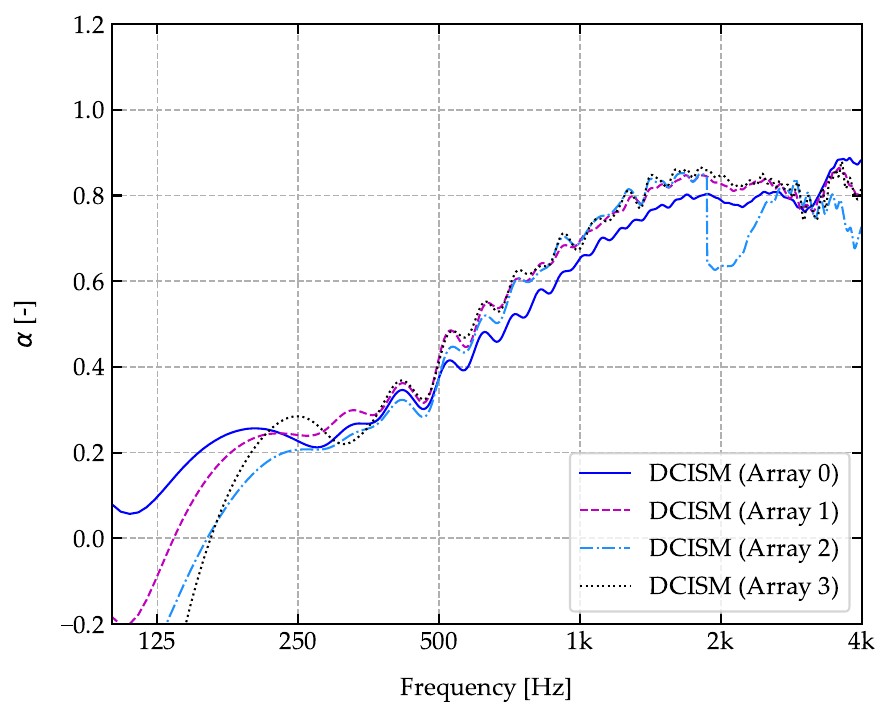}}
\subfloat[]{\label{fig:petmel_meas_arraycomp-b}
\includegraphics[width=0.49\linewidth]{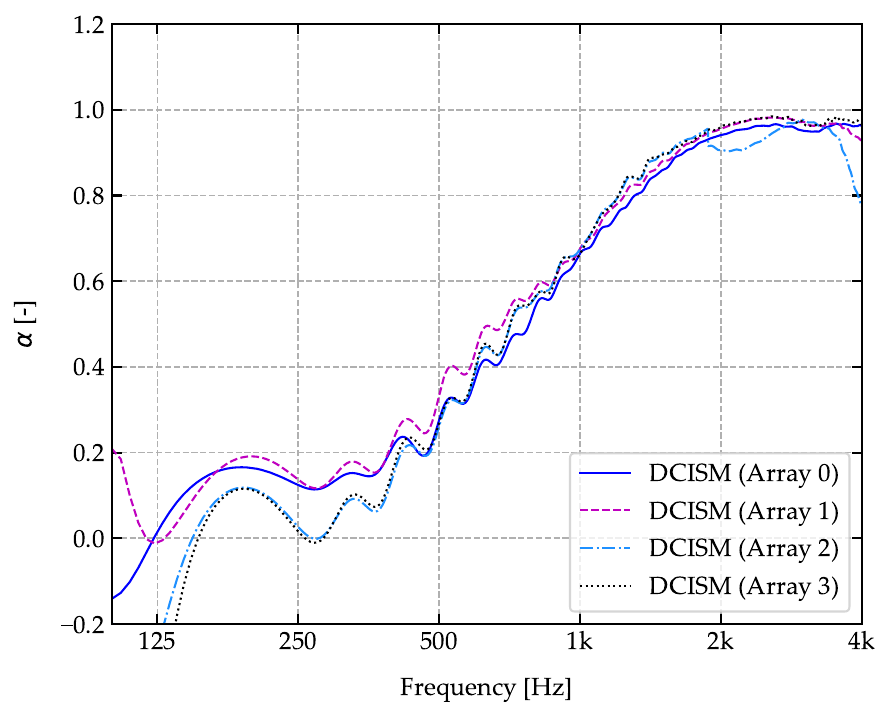}}
\caption{\co Sound absorption coefficient for the measurement of two finite porous absorbers obtained from Arrays~0--3 with the DCISM (GLE discretization); (a) PET porous absorber; and (b) Melamine porous absorber.}%
\label{fig:petmel_meas_arraycomp}
\end{figure}

\Figure{fig:petmel_meas_arraycomp} shows the predicted absorption coefficient for the PET and the Melamine porous absorbers computed with the DCISM (GLE discretization) for the data collected with Arrays~0--3. Notably, the variation across the arrays from 225~Hz to 1~kHz is relatively small compared to the simulations of \Secs~\ref{sec:results_inf} and \ref{sec:results_fin}. Again, one can observe the sudden variation in the absorption coefficient obtained from Array~2 near 2~kHz for both samples, consistent with the simulated data (see \Figs{}~\ref{fig:melamine_inf_alpha} and \ref{fig:melamine_fin_alpha-a} and attributed to the impact of the geometry of the array on the conditioning of matrix $\vecb{G}$.

\vspace{0.5cm}
It is also interesting to note that reliable estimations of the sound absorption coefficient are attainable with Array~3. This is attributed to exaggerating the effects of added noise in the simulations, which can be seen as a worst-case scenario.  Thus, smaller datasets can render reliable results, a feature of practical importance.  For example, when performing an extensive scan for an array such as Array~0, a smaller dataset can be scanned first and used to check if the measurements are accurate enough. Of course, this study did not aim to exhaust all possibilities of failure, and collecting more data is regarded as more secure.

\begin{figure}[ht!]
\begin{center}
\includegraphics[width=0.5\columnwidth]{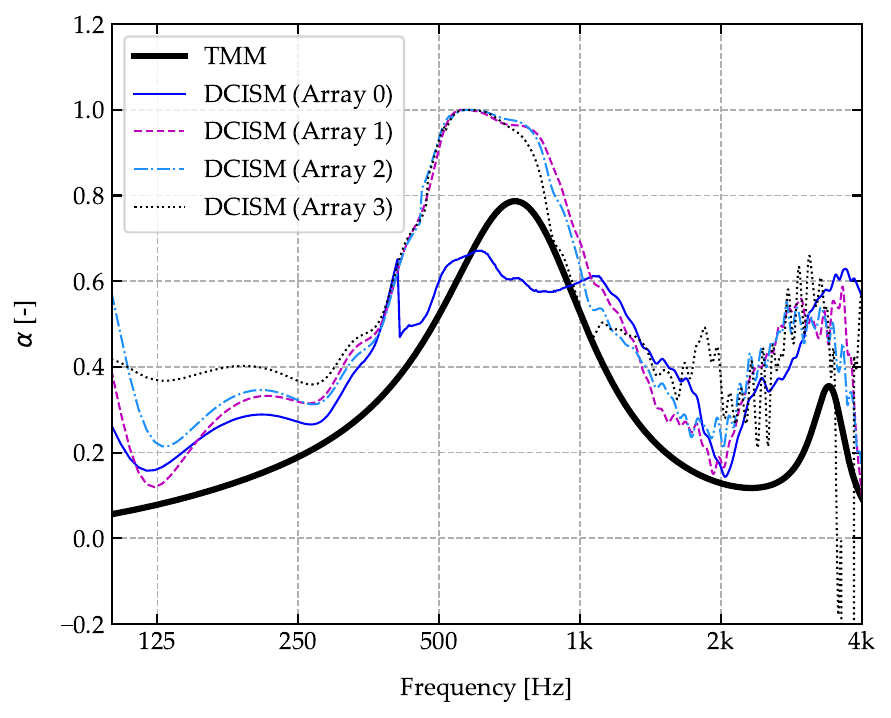}
\caption{\label{fig:perfplate_meas_alpha}{\co Sound absorption coefficient for the measurement of a finite slotted panel absorber obtained from Array~0--3 with the DCISM (GLE discretization); for reference, TMM simulation from Barbosa \myetal{} \cite{barbosacob}.}}
\end{center}
\end{figure}

\Figure{fig:perfplate_meas_alpha} displays the experimental absorption coefficient obtained from the DCISM for data collected by Arrays~0--3 for the Helmholtz absorber. The experimental data is compared to the TMM simulation in \Reference~\cite{barbosacob}. For this absorber, it was necessary to increase the number of Gauss points, $N$, in the GLE discretization to $N = 70$ while keeping the upper limit of integration the same as for the porous sample ($b = 20$). Thus, the sampling of the integration interval is discretized at a higher resolution. The increase of $N$ was necessary for the convergence of the results, and the value $N = 70$ was chosen after trial and error. A more formal analysis of the DCISM convergence is outside the scope of this research, but it can be pursued in the future. It is easy to note that the DCISM for Arrays~1, 2, and 3 render similar absorption coefficients. One can note the main absorption peak between 500--800~Hz, a dip in the absorption around 2~kHz, and another peak near 4~kHz. Overall, the measured absorption coefficient (Arrays~1--3) and the bandwidth of the main peak are overestimated relative to the TMM, which may be too simplistic for such a complex absorber. 

The absorption coefficient estimated with the DCISM for Array~0 does not agree with estimates made with the data of Arrays~1--3, being underestimated even relative to the TMM data up to the main peak of absorption. While the PET and the Melamine porous absorbers are placed directly on the floor of the measurement space, the Helmholtz absorber is made of a hard-wooden frame (see \Fig{fig:win_slotted_abs-a}). In contrast, there is experimental evidence that the \insitu{} measurement of porous materials with a hard frame around it render more edge diffraction effects than non-framed porous absorbers~\cite{Brandao2011a}. Thus, the edge diffraction effect may play a more important role for such a Helmholtz absorber, especially for a large aperture array, such as Array~0. The impact of edge diffraction on the \insitu{} measurement tends to be sample-dependent, which justifies the need for more research on the topic. 

\section{Conclusions} \label{sec:conclusions}


The research within this article proposes an inverse problem to measure the sound absorption coefficient of acoustical materials via \insitu{} measurements. The underlying mathematical model is based on discretizing an integral equation, mapping the measured sound pressure to a monopole source, an image source, and distributed monopoles along a line in the complex plane (DCISM). 
The proposed method underwent comparison with an inverse problem approach utilizing the source/image-source method (ISM). Incorporating additional monopoles into the sound field was advantageous, as it activated more than just the source/image source within the DCISM framework.
That led the proposed method to a significantly lower reconstruction error than the ISM. The research also demonstrates that the Gauss-Legendre quadrature discretization scheme favors the DCISM. The estimated absorption coefficient by the DCISM agrees with the one obtained by measuring the surface impedance at the sample's center for infinite and finite porous materials. The difference between spherical and \planewave{} incidence absorption coefficients is more significant for low-flow resistivity materials. The same is expected for smaller source-to-sample distances, which will be explored in future research. The article also demonstrates that practical measurements are possible even with cost-effective instrumentation and arrays containing just a few microphones. However, the measurement error increases at lower frequencies when small datasets are collected. Measurement errors may increase in the high-frequency range, depending on the geometry of a particular array. The experiments showed that the proposed technique is robust and that the effect of sample size is less prominent when compared to the simulations. More research on that aspect is necessary, as this feature is sample-dependent. For instance, consistent estimations of the absorption coefficient were possible for the measurement of the porous absorber with arrays of relatively large apertures, but for the resonant panel only if the array aperture was smaller than the sample. Thus, optimizing the array for constraints such as measurement apparatus compactness, the influence of noise, and the sample size effect on the reconstruction error are interesting topics for future research.

\appendix

\section{The Johnson-Champoux-Allard model} \label{sec:appendixA}

The Johnson-Champoux-Allard (JCA) model states that the characteristic density and the bulk modulus are given, respectively, by 
\begin{equation}
\label{eq:rho_jca}
\rho_{\text p}=\rho_{0} \ \alpha_{\infty}\left(1+\frac{\; \sigma \phi}{\; \ji \omega  \rho_{0} \, \alpha_{\infty}\;} \sqrt{1+\frac{\; 4 \ji \alpha_{\infty}^{2} \eta \rho_{0} \omega \;}{\sigma_{r}^{2} \Lambda^{2} \phi^{2}}}\right)
\end{equation}
\vspace{0.25em}
and
\vspace{0.25em}
\begin{equation}
\label{eq:k_jca}
\kappa=\frac{\gamma P_{0}}{\; \; \gamma-(\gamma-1)\left(1+\frac{8 \eta}{\, \ji \omega P_{r} \rho_{0} \Lambda^{'2} \,} \sqrt{1+\frac{\, \ji \omega \rho_{0} P_{r} \Lambda^{'2} \,}{16 \eta}}\right)^{\!\!-1} \;\;} \;,
\end{equation}
where $\alpha_{\infty}$, $\sigma$, $\phi$, $\Lambda$, and $\Lambda^{'}$ are given in \Tab{tab:samples}. The specific heat ratio is $\gamma=1.41$, $P_{0}=101325$~Pa is the atmospheric pressure, $P_{r}=0.71$ is the Prandtl number, $q_{0}=\eta / \sigma$ and $q_{0}^{\prime}=q_{0} \, \alpha_{\infty}$ are the viscous and thermal permeability, respectively. The air's dynamic viscosity is $\eta=1.81 \times 10^{-5}$~Pa$\cdot$s. Then, the complex \wavenumber{} is computed by $k_{\text{p}} = \omega \sqrt{\rho_{\text p}/\kappa}$, and the characteristic impedance of the porous layer is $Z_{\text{p}} = \sqrt{\kappa \rho_{\text p}}$.

\section{Helmholtz absorber model} \label{sec:appendixB}

The surface impedance of the Helmholtz absorber model is computed using the Transfer Matrix Method (TMM). The absorber comprises two layers: the slotted panel layer lies over a layer of porous material applied directly over rigid backing. Therefore, the total acoustic surface impedance of the absorber is given by
\begin{equation}
\label{eq:Zs_helm}
Z_{\text{s}} = Z_{\text{H}} - \ji Z_{\text{p}} \frac{k_{\text{p}}}{\; k_{\text{pz}} \;} \cot{(k_{\text{pz}} \, d)} \,,
\end{equation}
with $Z_{\text{H}}$ representing the acoustic surface impedance of the panel and the second term on the right-hand side of \Eq{eq:Zs_helm} representing the surface impedance of the porous material layer. The term $Z_{\text{H}}$ is computed (as in Barbosa \myetal{}~\cite{barbosacob}) by
\begin{equation}
\label{eq:Zh}
Z_{\text{H}}=\frac{\; \ji \omega \rho_{0} t \;}{\Theta} 
\left[1-\frac{\; \tanh \left(w \sqrt{\frac{\, \rho_{0} \omega \,}{4 \eta}} \sqrt{\ji}\right) \;}{w \,  \sqrt{\frac{\, \rho_{0} \omega \,}{4 \eta}} \sqrt{\ji}}\right]^{-1} + \,
\frac{\; 4 \sqrt{2 \rho_{0} \eta \omega} \;}{\Theta}-\Psi(\Theta) \,,
\end{equation}
where $t = 6$~mm is the panel thickness, and $w = 8$~mm is the panel width. Moreover, $\Theta=\left(S / S_{p}\right) = 0.126$ is the perforated ratio, with $S$ being the sum of the cross-sectional area of all slits, and $S_{p} = 0.6 \times 0.6$~m$^2$ being the total active area of the panel (borders excluded). In this research 
\begin{equation}
\Psi(\Theta) = \ji \, 0.936 \, \omega \, \rho_{0} \frac{\, w \, F(\Theta) \,}{2} \frac{\, \ln (\sin (\pi \, \Theta)) \,}{\Theta}
\end{equation}
and $F(\Theta)=1-1.25 \, \Theta$.

\section*{Acknowledgements} \label{sec:ack}

This work was supported by National Research Council of Brazil (CNPq - Conselho Nacional de Desenvolvimento Científico e Tecnológico) [grant number 402633/2021-0] and by Coordination of Superior Level Staff Improvement (CAPES - Coordenação de Aperfeiçoamento de
Pessoal de Nível Superior)  [grant number 88887.488185/2020-00]. The authors also thank Leandro Barbosa at Wave Consultoria, for providing two of the measured samples, data, and valuable discussions.


\bibliographystyle{elsarticle-num} 

%
%
\end{document}